\documentclass[11pt]{article}
\pdfoutput=1 
\usepackage{jheppub}
\usepackage{graphicx,verbatim}
\usepackage{psfrag, times}
\usepackage{upgreek}
\usepackage[T1]{fontenc}
\usepackage[vcentermath]{youngtab}

\usepackage{epsf}

\def\be{\begin{equation}}
\def\ee{\end{equation}}
\def\bea{\begin{eqnarray}}
\def\eea{\end{eqnarray}}
\def\bc{\begin{center}}
\def\ec{\end{center}}

\def\bR{{\mathbb{R}}}

\def\bQ{{\bf{Q}}}

\def\cC{{\mathcal{C}}}

\def\cF{{\mathcal{F}}}
\def\cM{{\mathcal{M}}}
\def\cN{{\mathcal{N}}}

\def\cS{{\mathcal{S}}}
\def\cT{{\mathcal{T}}}
\def\cI{{\mathcal{I}}}

\def\cL{{\mathcal{L}}}

\def\cV{{\mathcal{V}}}

\def\cZ{{\mathcal{Z}}}

\def\al{\alpha}

\def\nn{\nonumber}
\def\lam{\lambda}
\def\blam{\bar{\lambda}}

\def\bq{{\bf q}}

\def\tl{\tilde{l}}
\def\r2{{\sqrt{2}}}

\def\bZ{\bar{Z}}

\def\al{\alpha}
\def\bt{\beta}
\def\dal{{\dot{\alpha}}}
\def\dbt{{\dot{\beta}}}
\def\bxi{{\bar{\xi}}}

\def\sig{\sigma}
\def\bsig{\bar{\sigma}}
\def\tl{\tilde{l}}

\def\eps{\epsilon}

\def\btau{{\bf \tau}}
\def\ra{{\rm a}}
\def\rb{{\rm b}}
\def\bQ{{\bf Q}}
\def\cV{{\mathcal{V}}}
\def\bphi{\bar{\phi}}
\def\bpsi{\bar{\psi}}
\def\hI{\hat{I}}
\def\hJ{\hat{J}}
\def\hK{\hat{K}}
\def\ha{\hat{a}}
\def\hmu{\hat{\mu}}

\def\bZ{{\mathbb{Z}}}
\def\bY{{\mathbb{Y}}}
\def\bj{{\bf{j}}}
\def\bX{{\bf{X}}}
\def\rM{{\rm{M}}}
\def\br{{\bf{r}}}
\def\bB{{\mathbb{B}}}
\def\rr{{\rm{r}}}
\def\rs{{\rm{s}}}
\def\hm{\hat{m}}

\def\bmu{{\upmu}}
\def\bnu{{\upnu}}
\def\bfZ{{\bf{Z}}}
\def\oY{\overline{Y}}

\def\bt {\beta}
\def\ra {{\rm a}}
\def\rb {{\rm b}}

\def\nb {\nonumber}
\def\pd {\prod}
\def\fc {\frac}

\def\rla{{l_{\ra}}}
\def\rlb{{l_{\rb}}}

\begin{document}
\title{\Large On Higgs Branch Localization of Seiberg-Witten Theories on Ellipsoid.}
\author[]{Heng-Yu Chen${}^{1}$ and Tsung-Hsuan Tsai${}^{1}$}
\affiliation{$^1$Department of Physics and Center for Theoretical Sciences, \\
National Taiwan University, Taipei 10617, Taiwan}
\emailAdd{ heng.yu.chen@phys.ntu.edu.tw, r01222053@ntu.edu.tw} 
\vspace{2cm}
\abstract
{In this note, we consider so-called ``Higgs Branch Localization'' for four dimensional $\cN=2$ field theories on 4d ellipsoid.
We find a new set of saddle point equations arising from additional Higgs branch deformation term, 
whose solutions include both Higgs branch and BPS instanton-vortex mixed configurations.
By evaluating the contour integral, we also demonstrate the ellipsoid partition almost factorizes into purely $b$ and $b^{-1}$ dependent parts, using SQCD as an explicit example.
We identify various factorized parts with the ellipsoid partition function of two dimensional $\cN=(2,2)$ SQCDA, which is precisely the vortex world volume theory.
We also give physical interpretation for the non-factorizable parts and discuss future directions.
 }

\maketitle

\section{Introduction}\label{Sec:Intro}
\paragraph{}
Beginning with a supersymmetric field theory $\cT$ in flat space and putting it on a curved manifold $\cM$,
the standard supersymmetric localization principle states we need to construct certain fermionic supercharge 
$\bQ$ consisting typically of global symmetry, e. g. $R$-symmetry and isometry of $\cM$,
the supersymmetry transformations on $\cM$ are then parameterized by the spinors satisfying 
the Killing equations with background fluxes turned on.
We can next construct $\bQ$-closed functional $\cS$ and deform it by a  $\bQ$-exact functional satisfying $\bQ^2 \cV =0$, 
i. e. $\cS \to \cS+t\bQ \cV$, as we set the deformation parameter $t \to +\infty$, the partition function $\cZ_{\cM}$ localizes along the saddle point loci $\bQ \cV =0$.
The choice of deformation functional $\bQ\cV$ can be somewhat arbitrary, 
other than the obvious requirement of positive semi-definiteness and existence of SUSY-preserving saddle point loci, 
one can further deform $\bQ\cV$ to $\bQ[\cV+\Delta \cV]$ while leaving $\cZ_{\cM}$ invariant, as it is only sensitive to the deformations in $\bQ$-cohomology.
However the new deformation can lead to new distinct saddle point loci, some correspond to different regions in the moduli space of $\cT$, often they can also be regarded as the defining equations of BPS solitons defined on some sub-manifold $\cC\subset \cM$. 
This gives rise to the factorization properties of $\cZ_{\cM}$ first noticed in \cite{Pasquetti:2011fj} for $\cM$ being 3d ellipsoid (see also \cite{Beem:2012mb}), 
and later realized in \cite{Benini:2012ui}, \cite{Doroud:2012xw} for $\cM=S^2$ that this can be understood as the consequence of so-called ``Higgs branch'' localization.
The essence is that we add to the usual Coulomb branch deformation $\cV_{\rm Coulomb}$ another deformation functional $\cV_{\rm Higgs}$,
such that the new saddle point loci now include part of Higgs branch and non-trivial gauge configurations, which are curved space analogue of dynamical vortices on $\cM$. 
Higgs branch localization has been applied to supersymmetric gauge theories on $S^1\times S^2$ and 3d ellipsoid \cite{Benini:2013yva}, \cite{Fujitsuka:2013fga} and $S^1\times S^3$ \cite{Peelaers:2014ima}.
In this note, we would like to initiate Higgs branch localization of four dimensional $\cN=2$ supersymmetric field theory or ``Seiberg-Witten theory'' on $S_{b^2}^4$\footnote{We will made precise our definition of deformation parameter in next section.}, 
whose partition function has been computed in a beautiful work \cite{Hama:2012bg} using the more common Coulomb branch localization,
however it remains well-motivated to perform such a task for the following reasons.
\paragraph{}
First, in contrast with the localization on other manifolds, even before adding $\cV_{\rm Higgs}$, Coulomb branch localization on $S^4$ or $S_{b^2}^4$ already involves singular gauge configurations, i. e. instantons and anti-instantons  \cite{Hama:2012bg}, \cite{Pestun:2007rz}.
The addition of $\cV_{\rm Higgs}$ modifies these into non-trivial vortex-instanton mixed configurations as shown in Section \ref{Sec:HiggsLoc}, and the integrated partition function predicts the equivariant volume of their yet to be studied moduli space. 
Second, as we will show explicitly in Section \ref{Sec:4dEvaluation} using 4d $\cN=2$ SQCD as an example, 
its partition function on $S_{b^2}^4$ despite containing both perturbative and non-perturbative contributions, 
again exhibits almost factorizable structures after explicit contour integration (See equation (\ref{Z: 4dSQCD-res})). 
The factorized parts can be identified with the $S_b^2$ (or $S^2_{1/b}$) partition function of 2d $\cN=(2,2)$ SQCD plus an adjoint chiral multiplet (denoted ``SQCDA''), which is precisely the world volume theory of dynamical vortices. This is in similar spirit  of  ``bootstrapping superconformal indices'' \cite{Gaiotto:2012xa}, \cite{Gaiotto:2014ina}, 
where the residues of a pole in fugacity for certain gauge/global symmetry can be interpreted as the index without this symmetry but with additional co-dimension two defects, here we extend this to $\cZ_{S^4_{b^2}}$ which contains additional non-perturbative instanton contributions, it would be interesting to replace the step of evaluating residue by the action of certain shift operator.
Finally, it is well-known that the partition function of $\cN=2$ SCFTs on $S^4$ or $S^4_{b^2}$ plays the pivotal role in the conjecture of \cite{Alday:2009aq}, 
and the co-dimension two surface defects which include IR limit of the dynamical vortices studied here, are a class of interesting observables labeled by different representations of the global symmetry group. The current setup enables us to construct other variety of 2d surface defects from the Higgs branch localization of appropriate 4d gauge theories.
\paragraph{}
This notes is organized as follows. 
In Section \ref{Sec:HiggsLoc}, we consider the Higgs branch localization of $\cN=2$ supersymmetric field theories on 4d ellipsoid $S_{b^2}^4$, 
and derive the new set of saddle point equations. 
In Section \ref{Sec:4dEvaluation}, we explicitly evaluate the contour integral of $\cZ_{S_{b^2}^4}$ for 4d $\cN=2$ SQCD, 
demonstrate its almost factorizability from the residues and identify various components with the contributions from the new saddle points solutions.
In Section \ref{Sec:2dEvaluation}, we compute the $S_b^2$ partition function for 2d $\cN=(2,2)$ SQCDA which is the vortex world volume theory,
and match it precisely with the suitable components from 4d residues. We leave some technical details in  Appendices \ref{App:Spinors} and \ref{App:SUSYTrans}.
\paragraph{}
Note Added: We are grateful to the authors of \cite{Pan:2015hza}, whose recent publication and communication with us help us to correct the mistakes and clarify the confusion in our previous version.

\section{Higgs Branch Localization on $S_{b^2}^4$}\label{Sec:HiggsLoc}
\subsection{Derivation of new saddle point equations}
\paragraph{}
The four dimensional ellipsoid $S_{b^2}^4$ can be described by its embedding equation:
\be\label{EmbedEqn}
\frac{x_0^2}{r^2}+\frac{x_1^2+x_2^2}{l^2} + \frac{x^2_3+x_4^2}{\tl^2} = 1
\ee
where $b^2 = \frac{l}{\tilde{l}}$ is defined to be the deformation parameter. 
This manifold possess $U(1)\times U(1)$ isometry rotating $(x^1, x^2)$ and $(x^3, x^4)$ planes with the angular coordinates denoted by $(\varphi, \chi)$,
its Killing vector is given by:
\be\label{KillingVector}
v^m \partial_m = \frac{1}{l} \partial_{\varphi} + \frac{1}{\tl} \partial_{\chi}.
\ee
The fixed points of $U(1)\times U(1)$ rotation are located at $x_0 = \pm r$, which correspond to north and south poles of $S^4_{b^2}$, 
while in the $b^2 \to 0$ or $b^{-2} \to 0$ limit, only $\varphi$ or $\chi$ rotation manifests.
We can explicitly satisfy the embedding equation (\ref{EmbedEqn}) by the following embedding coordinates and their associated vielbein one forms:
\bea\label{coords}
&&x_0 = r\cos\rho, \quad \quad \quad \quad \quad ~ E^a=E^a_m dx^m,\nn\\ 
&&x_1= l\sin\rho\cos\theta\cos\varphi, \quad E^1=l\sin\rho\cos\theta d\varphi,\nn \\
&&x_2=l\sin\rho\cos\theta\sin\varphi,\quad E^2=\tl\sin\rho\sin\theta d\chi, \nn \\
&&x_3= \tl\sin\rho\sin\theta\cos\chi,\quad E^3= f\sin\rho d\theta+h d\rho,\nn\\
&&x_4=\tl\sin\rho\sin\theta\sin\chi,\quad E^4= g d\rho,\nn
\eea
where we also defined the following combinations of coordinates:
\be
f(\theta)=\sqrt{l^2\sin^2\theta + \tl^2\cos^2\theta}, \quad g(\theta, \rho) = \sqrt{r^2\sin^2\rho + \frac{l^2 \tl^2}{f^2(\theta)}\cos^2\rho}, \quad h(\theta,\rho)=\frac{\tl^2-l^2}{f(\theta)} \cos\rho\sin\theta\cos\theta.
\ee
In terms of (\ref{coords}) the the north or south pole of $S_{b^2}^4$ respectively corresponds to $\rho=0$ or $\rho=\pi$, 
near each of these two points say $x_0 \simeq r$, the remaining coordinates become:
\be
x_1 \simeq  l\rho\cos\theta\cos\varphi, ~~ x_2 \simeq  l\rho\cos\theta\sin\varphi,~ x_3 \simeq  \tl\rho\sin\theta\cos\chi, ~~ x_4 \simeq  \tl\rho\sin\theta\sin\chi,
\ee
which are precisely the rescaled polar coordinates of ${\bR}^4$ for $l\neq \tl$. It was demonstrated in \cite{Hama:2012bg} that the Killing spinor hence the supercharges on $S_{b^2}^4$ reduce in this further limit to those for four dimensional $\Omega$-background \cite{Nekrasov:2002qd} \cite{Nekrasov:2003rj}, it is therefore natural to identify the $\Omega$-deformation parameters in this limit as: $\eps_1= \frac{1}{l}, \eps_2 =\frac{1}{\tl}$. While $b^2 \to 0$ or $b^{-2} \to 0$ now becomes the limit taken in \cite{Nekrasov:2009rc} when we consider the quantization of integrable systems using exact gauge theory partition functions.
Moreover when $\theta = 0$ (or $\theta = \frac{\pi}{2}$), the non-vanishing $(x_0, x_1, x_2)$ and their vielbeins (or $(x_0, x_3, x_4)$) reduce to those 
for a deformed $S^2$, where its deformation parameter is $b=\frac{l}{r}$ (or $\frac{1}{b}=\frac{\tl}{r}$), we therefore denote it as $S_{b}^2$ (or $S^2_{1/b}$). 
As demonstrated in \cite{Drukker:2010jp}, co-dimension two BPS surface defects can wrap on these two deformed two spheres,  
the north and south poles of $S^2_{b}$ and $S^2_{1/b}$ precisely coincide with those of $S^4_{b^2}$ hence each other, 
this simple fact has interesting consequence when identifying the non-perturbative 4d instantons and 2d world sheet instantons later.
\paragraph{}
Let us next consider putting on $S_{b^2}^4$ a class of field theories with $\cN=2$ supersymmetry on $\bR^4$, 
the supersymmetry transformations on various field contents are characterized by a pair of $SU(2)_R$ doublet chiral and anti-chiral Killing spinors $\xi \equiv (\xi_{\al A},\bxi^{\dal}_{~A})$, satisfying the equations:
\bea
&&D_m\xi_{A}  = \partial_m \xi_{A} + \frac{1}{4}\Omega_m^{ab}\sigma_{ab}\xi_{A} + i \xi_{B} (V_m)^B_{~A} = - (T^{kl}\sig_{kl} \sig_m +\sig_m \bar{S}^{kl}\bsig_{kl} )\bxi_A\label{KS1}
\\
&&D_m\bxi_{A} =\partial_m \bxi_{A} + \frac{1}{4}\Omega_m^{ab}\bsig_{ab}\bxi_{A} + i \bxi_{B} (V_m)^B_{~A} = - (\bar{T}^{kl}\bsig_{kl}\bsig_m + \bsig_m {S}^{kl}\sig_{kl}) \xi_A,\label{KS2}
\eea 
where $A = 1,2 $ is the $SU(2)_R$ index, and we have contracted $\al, \dal=1,2 $ which are respectively
chiral spinorial and anti-chiral spinorial indices\footnote{Our index notations and contraction conventions 
are summarized in the beginning of Appendix \ref{App:Spinors}.},
and the non-vanishing components of the spinor connections $\Omega_{m}^{ab}$.
Here we have also introduced following auxiliary background fields: $T_{kl}$ and $\bar{T}_{kl}$ are rank two space-time tensors,
and $(V_m)^B_{~A}$ are background gauge fields for $SU(2)_R$ symmetry, all of them vanish in the round $S^4$ limit $r=l=\tl$ (i. e. $b^2=1$). 
While the anti-symmetric tensors $S^{kl}$ and $\bar{S}^{kl}$ can be regarded as arising from the curvature coupling with the fermions. 
We summarize the explicit expressions for $T_{kl},\bar{T}_{kl}, S_{kl}, \bar{S}_{kl}$ in the appendix \ref{App:SUSYTrans}.
In addition, for supersymmetry algebra to close off-shell, $(\xi_A,\bxi_A)$ also need to satisfy the following auxiliary equations
:
\bea
&&\sigma^m\bsig^n D_m D_n\xi_A+4 D_l T_{mn}\sig^{mn}\sig^l\bxi_A = {\bf{M}} \xi_A,\label{AXE1}\\
&&\bsig^m\sig^n D_m D_n \bxi_A+4 D_l \bar{T}_{mn}\bsig^{mn}\bsig^l\xi_A = {\bf{M}} \bxi_A.\label{AXE2}
\eea
where ${\bf{M}}$ is a background scalar field relating to the curvature of $S_{b^2}^4$.
The origin of these auxiliary fields $T_{mn}, \bar{T}_{mn}, V_m$ and ${\bf M}$ can be traced back to the supergravity multiplets when we couple the field theories on ${\mathbb R}^4$ to off-shell supergravity.
The important insight in \cite{Hama:2012bg} was that we can consider the Killing spinors for round $S^4$:
\be
\xi_{\alpha A} = \frac{1}{2}\sin\frac{\rho}{2}
 \begin{pmatrix} 
e^{\frac{i}{2}(\chi+\varphi-\theta)} & e^{-\frac{i}{2}(\chi+\varphi+\theta)} \\
-e^{\frac{i}{2}(\chi+\varphi+\theta)} & e^{-\frac{i}{2}(\chi+\varphi-\theta)}
\end{pmatrix},\quad
\bxi^{\dal}_{~A} = \frac{i}{2}\cos\frac{\rho}{2}
 \begin{pmatrix} 
e^{\frac{i}{2}(\chi+\varphi-\theta)} & -e^{-\frac{i}{2}(\chi+\varphi+\theta)} \\
-e^{\frac{i}{2}(\chi+\varphi+\theta)} & -e^{-\frac{i}{2}(\chi+\varphi-\theta)}
\end{pmatrix}.
\ee
These spinors can remain the solutions to  (\ref{KS1}), (\ref{KS2}), (\ref{AXE1}) and (\ref{AXE2}),
provided the aforementioned auxiliary fields take appropriate forms and they were solved explicitly in \cite{Hama:2012bg},
we summarize them in the appendix \ref{App:SUSYTrans}.
Moreover it is necessary for the Killing spinors to satisfy the following reality condition:
\be\label{KSReality}
(\xi_{\al A})^{\dag} = \xi^{A\al} = \eps^{\al\bt}\eps^{AB}\xi_{\bt B}, \quad (\bxi_{\dal A})^{\dag} = \bxi^{A\dal} = \eps^{\dal\dbt}\eps^{AB} \bxi_{\dbt B},
\ee
where $\eps^{AB}$ and $\eps_{AB}$ are the anti-symmetric $SU(2)_R$ invariant tensors. 
Notice that we can now use $(\xi_A, \bxi_A)$ to express the Killing vector $v^m$ of $S_{b^2}^4$ as $v^m = 2\bxi^{A}\bsig^m\xi_A$, 
and they are annihilated by $\bQ^2$, i. e. $\bQ^2 \xi_A = \bQ^2\bxi_A = 0$ where $\bQ$ is the supercharge used for the localization computation.
We can now parameterize the supersymmetry transformations of vector and hyper multiplets in terms of $(\xi_A, \bxi_A)$, 
also construct the supersymmetric invariant Lagrangians, these are summarized in Appendix \ref{App:SUSYTrans}.   
\paragraph{}
Focusing on the Higgs branch localization on $S_{b^2}^4$,
first we should note that for a given deformation term $\bQ {\cV}$, among all of its saddle point loci, 
the only non-vanishing contributions to the path integral come from the ones which coincide with the
supersymmetric field configurations specified by:
\be
\bQ {\bf \Psi} = 0, \quad {\rm for~all~fermions~{\bf \Psi}.} 
\ee
This naturally leads us to add to the supersymmetric Lagrangian, an additional manifestly positive semi-definite deformation term constructed from the vector multiplet \cite{Hama:2012bg}:
\be\label{VecDeform1}
\cI_{\rm vec.} = {\rm Tr} \left[(\bQ \lam_{\al A})^{\dag}(\bQ\lam_{\al A})+(\bQ\bar{\lam}^{\dal}_{~A})^{\dag}(\bQ\bar{\lam}^{\dal}_{~A})\right].
\ee 
Here $(\lam_{\al A}, \bar{\lam}^{\dal}_{~A})$ are a pair of chiral and anti-chiral gauginos transforming in the adjoint representation of the gauge group and their supersymmetric transformations are given in (\ref{GauginoTrans1}) and (\ref{GauginoTrans2}), the trace here is taken over gauge indices.
For our later purpose, we can explicitly express (\ref{VecDeform1}) into the following component form:
\bea\label{BosVecDeform}
\cI_{\rm vec.}\mid_{\rm Bose} &=& \sin^2\frac{\rho}{2} {\rm Tr}\left[(W_{mn}^{-})^2 -\frac{1}{2}(D_{AB}-(\phi+\bphi)w_{AB})(D^{AB}-(\phi+\bphi)w^{AB}) \right]\nn\\
&+&   \cos^2\frac{\rho}{2} {\rm Tr}\left[(W_{mn}^{+})^2 -\frac{1}{2}(D_{AB}-(\phi+\bphi)w_{AB})(D^{AB}-(\phi+\bphi)w^{AB}) \right]\nn\\
&+&{\rm Tr}\left[-D_m (\phi+\bphi)D^m(\phi+\bphi)+4[\phi,\bphi]^2+\frac{ (v^m D_m(\phi-\bphi))^2  }{4\sin^2\frac{\rho}{2}\cos^2\frac{\rho}{2}}\right],
\eea
where $SU(2)_R$ tensor $w^{AB}$ is given in (\ref{Def:wAB}),
we have also rewritten $\cI_{\rm vec.}$ in slightly different form from \cite{Hama:2012bg} and defined:
\bea
&&W^{-}_{mn} = F_{mn}^{-}-4(\phi-\bphi)(T_{mn}+S_{mn})+\frac{v_{[m}D_{n]}(\phi-\bphi)}{\sin^2\frac{\rho}{2}},\\
&&W^+_{mn} =  F_{mn}^{+}-4(\phi-\bphi)(T_{mn}+S_{mn}) +\frac{v_{[m}D_{n]}(\phi-\bphi)}{\cos^2\frac{\rho}{2}}.
\eea
Notice that the deformation term for the vector multiplet is manifestly positive semi-definite with respect to the reality conditions given in (\ref{RealityCond-Vec1}) and (\ref{RealityCond-Vec2}).
\paragraph{}
Now to perform Higgs branch localization we can add to $\cI_{\rm vec.}$ another $\bQ$-exact deformation term given by:
\bea\label{HiggsDeform}
{\cI}_{\rm Higgs} &=&  2\bQ{\rm Tr} \left[ (\lambda_{\al A})^{\dag}H_{\al A}+(\bar{\lambda}_{~A}^{\dal})^{\dag}\bar{H}^{\dal}_{~A}\right].
\eea
Here $H_{\alpha A}$ and $\bar{H}^{\dal}_{~A}$ are:
\be\label{Def:HA}
H_{\al A} = i H(\phi, q) ( \tau_{\theta}^1)_\al^{~\bt}\xi_{\bt A}, \quad \bar{H}^{\dal}_{~A} = iH(\phi, q)(\tau^1_{\theta})^{\dal}_{~\dbt}\bar{\xi}^{\dbt}_{~A},
\ee
\be
i\tau_{\theta}^1=i(\cos\theta\tau_1+\sin\theta\tau_2) = (\cos\theta \sig_{41}+\sin\theta\sig_{42})_{\al}^{~\bt} = -(\cos\theta\bsig_{41}+\sin\theta\bsig_{42})^{\dal}_{~\dbt}
\ee
and $H(\phi, q)$ is a hermitian functional satisfying $H(\phi, q) = H(\phi, q)^{\dag}$, which can contain the scalar fields in both vector multiplet $\phi, \bphi$ and hypermultiplet $q_A$, also FI-parameter $\zeta$.  
Also notice that $\tau_\theta^1 \propto v^m \sig_m$ such that it can be directly constructed from the Killing spinors and preserve the isometry of $S_{b^2}^4$.
To find the saddle point loci of the combined (\ref{VecDeform1}) and (\ref{HiggsDeform}), we first expand out the bosonic parts of (\ref{HiggsDeform}) into:
\bea\label{BosonicDef.}
\cI_{\rm Higgs}\mid_{\rm Bose} 
&=&\sin^2\frac{\rho}{2}{\rm Tr}\left[H(q, \phi)\left\{2(\cos\theta({\cF}_{23}-{\cF}_{14})+\sin\theta({\cF}_{13}-{\cF}_{24}))
+i(D^{AB}-(\phi+\bphi)w^{AB})(\tau_1)_{AB})\right\}\right]\nn\\
&+& \cos^2\frac{\rho}{2}{\rm Tr}\left[H(q,\phi)\left\{2(\cos\theta(\bar{\cF}_{23}+\bar{\cF}_{14})+\sin\theta(\bar{\cF}_{13}+\bar{\cF}_{24}))
+i(D^{AB}-(\phi+\bphi)w^{AB})(\tau_1)_{AB})\right\}\right]
\nn\\
&+&4\sin\frac{\rho}{2}\cos\frac{\rho}{2}{\rm Tr}[ D_4 (\bphi-\phi)].
\eea
where $\mathcal{F}_{mn} = F_{mn}-4(\phi-\bphi)(T_{mn}+S_{mn})$, $\bar{\cF}_{mn}=F_{mn}-4(\phi-\bphi)(\bar{T}_{mn}+\bar{S}_{mn})$ and $\tau_1$ is Pauli matrix.
We notice that the combined deformations $\cI_{\rm vec.}+\cI_{\rm Higgs}$ is no longer positive semi-definite with respect to the previous reality conditions in \cite{Hama:2012bg}, 
instead we need to further relax the reality condition for the auxiliary field $(D_{AB})^{\dag} =- D^{AB}$ given in (\ref{RealityCond-Vec2}),  
and deform the integration contour so that $D_{12} = D_{21}$ components can now also pick up imaginary values, the resultant D-term constraint becomes:
\be\label{Dterm-constraint}
D_{AB} = (\phi+\bphi) w_{AB} + i H(q, \phi)(\tau_1)_{AB}.
\ee
Now we can readily write the components and complete the squares with the field strength terms into:
\bea\label{IvecHiggs-Bosonic}
&&\cI_{\rm vec.}+\cI_{\rm Higgs}\mid_{\rm Bose} \nn\\
&&= \sin^2\frac{\rho}{2}{\rm Tr}\left[(H(q,\phi)-\cos\theta(F_{32}-F_{41})-\sin\theta(F_{31}-F_{42}))^2\right]\nn\\
&&+\sin^2\frac{\rho}{2}\left[(\sin\theta(F_{32}-F_{41})-\cos\theta(F_{31}-F_{42}))^2+(F_{12}-F_{34})^2\right]\nn\\
&&+\cos^2\frac{\rho}{2}{\rm Tr}\left[(H(q,\phi)-\cos\theta(F_{32}+F_{41})-\sin\theta(F_{31}+F_{42}))^2\right]\nn\\
&&+\cos^2\frac{\rho}{2}{\rm Tr}\left[(\sin\theta(F_{32}+F_{41})-\cos\theta(F_{31}+F_{42}))^2+(F_{12}+F_{34})^2\right]\nn\\
&&-{\rm Tr}[D_m(\phi+\bphi)D^m(\phi+\bphi)-4[\phi,\bphi]^2] + {(\phi-\bphi)~{\rm dependent~terms}}.
\eea
Here we have collected and abbreviated the remaining terms which vanish identically when $(\phi-\bphi)=0$, 
as we expect non-trivial gauge configurations arising from the saddle point solutions would still need to obey this Coulomb branch-like condition.
\paragraph{}
Clearly we have hypermultiplets here and we also need to include the deformation terms for them to ensure saddle point solutions coincide with their supersymmetric loci,  a natural choice is given by
\footnote{An alternative choice of deformation was given in \cite{Hama:2012bg}, where the authors simply noticed that the hypermultiplet action is also $\bQ$-exact, i. e. $\cL^{\rm hyp.} = \bQ \cV^{\rm hyp.}$, however it can share the same saddle point loci as (\ref{HypDeform1}).}
\be\label{HypDeform1}
\cI_{\rm hyp.} = \frac{1}{4}{\rm Tr} [(\bQ {\psi}_{\al \hI})^{\dag}(\bQ{\psi}_{\al \hI})+
 (\bQ \bar{\psi}^{\dal}_{~ \hI})^{\dag}(\bQ\bar{\psi}^{\dal}_{~\hI})]
 \ee
where $\bQ\psi_{\al\hI}, \bQ\bpsi^{\dal}_{~\hI}$ are supersymmetric variations of the fermions in hypermultiplets given explicitly in (\ref{HyperTrans2}) and (\ref{HyperTrans3}). 
The corresponding saddle point equations are then given by:
\bea
&&\bQ \psi^{\hI} = 2(\sig^n \bxi_A) D_n q^{A\hI}-4[(S_{kl}\sig^{kl})+i(\bphi+\bar{\mu})]\xi_A q^{A\hI}+2\cot\frac{\rho}{2}\xi_A F^{A\hI}=0, \label{HyperSaddle1} \\
&&\bQ \bpsi^{\hI} = 2(\bsig^n \xi_A)D_n q^{A\hI}-4[(\bar{S}_{kl}\bsig^{kl})+i(\phi+{\mu})]\bxi_A q^{A\hI}-2\tan\frac{\rho}{2}\bxi_A F^{A\hI}=0. \label{HyperSaddle2}
\eea  
where the covariant derivative $D_n q^{A\hI}$ is defined in (\ref{Def:Dq}).
Here we have also included the complex mass parameters $(\mu, \bar{\mu})$.
It is important to note that $q^{A\hI}$ also couples to non-vanishing background $SU(2)_R$ gauge field $V_n$.
Here we can also work out the explicit bosonic component form of $\cI_{\rm hyp.}$
\footnote{We would like to stress that while we take $\mu$ to be complex for the time-being, as we will demonstrate later, 
to satisfy the BPS equation however it is necessary to impose reality condition, as in the case for scalar $\phi$.}
:
\bea\label{IHyp:Bosonic}
\cI_{\rm hyp.}\mid_{\rm Bose} &=& {\rm Tr}\left[-\frac{1}{2}F^{A\hI}F_{\hI A}-i\sin\frac{\rho}{2}\cos\frac{\rho}{2}[(\phi+\mu)-(\bphi+\bar{\mu})](F^{A\hI}q_{\hI A}+q^{A\hI}F_{\hI A})\right]\nn\\
&+&\frac{1}{2}{\rm Tr}\left[ D^m q^{A\hI}D_{m}q_{\hI A}+{\mathbb M} q^{A\hI} q_{\hI A}- q^{\hI A}  \bxi_A \bar{\sig}^{mn}\bxi_B [D_m, D_n] q^{B}_{~\hI} 
+ q^{\hI A}  \xi_A\sig^{mn}\xi_B[D_m, D_n]q^{B}_{~\hI} \right] \nn\\
&-&2i{\rm Tr}\left[((\phi+\mu)-(\bphi+\bar{\mu}))\left[q^{\hI A}\xi_A\sig^n\bxi_B D_n q^{B}_{~\hI}+D_n q^{\hI A} \xi_A\sig^n\bxi_B q^{B}_{~\hI}+q^{\hI A}\cM_{AB} q^{B}_{~\hI}\right]\right]\nn\\
&+&2{\rm Tr}\left[ D_nq^{\hI A}{\mathbb S}_{AB}^n q^{B}_{~\hI}  -  q^{\hI A}{\mathbb S}^{n}_{AB} D_{n}q^{B}_{~\hI} + q^{\hI A}\Xi_{AB}^nD_{n} q^B_{~\hI} \right]
\eea
where we have defined the following quantities in the expression above:
\bea
&&{\mathbb M} = \frac{(f+g)^2+h^2}{4f^2 g^2}-4(\phi+\mu)(\bphi+\bar{\mu}),\label{MassMatrix1}\\
&&{\mathcal M}_{AB} = \frac{\cos\rho}{2fg} \begin{pmatrix}
(f+g) & ie^{-i(\chi+\varphi)}h\\
-ie^{i(\chi+\varphi)}h& -(f+g)
\end{pmatrix},\label{MassMatrix2}\\
&& {\mathbb S}_{AB}^n = \bxi_A \bsig^n (S_{kl} \sig^{kj})\xi_B-\xi_A\sig^n (\bar{S}_{kl}\bar{\sig}_{kl})\bxi_B,\label{SAB}\\
&&\Xi_{AB}^n = D_m\bxi_A\bsig^{mn}\bxi_B + \bxi_A \bsig^{mn}D_m \bxi_B-D_m\xi_A\sig^{mn}\xi_B-\xi_A\sig^{mn}D_m\xi_B.\label{XiAB}
\eea
We can now integrate out the auxiliary field $F^{\hI A}$ in the first line of (\ref{IHyp:Bosonic}) to impose the constraint:
\be\label{F-constraint}
F_{\hI A} = -i\sin\rho [(\phi+\mu)-(\bphi+\bar{\mu})] q_{\hI A}
\ee
and replace the $F_{\hI A}$ dependent terms with $-\frac{1}{2} \sin^2\rho [(\phi+\mu)-(\bphi+\bar{\mu})]^2 q^{A\hI}q_{\hI A}$.
Notice that while we can in principle compute explicitly the matrices ${\mathbb S}_{AB}^n$ and $\Xi_{AB}$ using (\ref{KS1}) and (\ref{KS2}), 
however it is worth noting that each term contains both $\xi_A$ and $\bxi_A$ and vanish identically at $\rho=0$ and $\rho=\pi$. 
This become useful when we study the possible saddle point solutions.

\subsection{New saddle point solutions}\label{SubSec:New Saddle}
\paragraph{}
To solve for the saddle point solutions giving:
\be\label{SaddlePtEqn: General}
\cI_{\rm vec.}+\cI_{\rm Higgs}+\cI_{\rm hyp.}\mid_{\rm Bose} = 0 
\ee
let us begin by briefly recalling the simplest case where $H(\phi, q)=0$ identically \cite{Hama:2012bg} and we also set $\mu=\bar{\mu}=0$ for the time being, 
there can be two distinct classes of solutions:
\paragraph{}
{\bf Coulomb branch-like solutions:}
These are smooth solutions existing everywhere on $S^4_{b^2}$, 
up to gauge transformation they are given by:
\be\label{Saddle:Coulomb}
A_n = F_{mn}=0, ~~ \phi=\bphi =  -i \frac{a_0}{2}, ~~ D_{AB} = -i a_0 w_{AB}, ~~ q^{A\hI}= F^{A\hI} =0
\ee
where $a_0$ is a constant real valued matrix taking values in the Cartan of the gauge group;  
$w_{AB} = w_{BA}$ is a rank two symmetric $SU(2)_R$ tensor defined in terms of the Killing spinors and auxiliary fields, it is explicitly given in (\ref{Def:wAB}).
The name ``Coulomb branch'' here can be attributed to the expectation values for the hypermultiplet scalar $q_{A \hI}$ is set to vanish by the finite curvature of $S_{b^2}^4$ which cannot be canceled by tuning the vector scalar expectation value $\phi=-\frac{i}{2}a_0\in -i{\mathbb R}$.
This also implies that for the Higgs branch to exist, we need to allow for $a_0$ to be complex as we do when performing the contour integration, 
and introduce additional parameters such as complex masses in order to cancel the positive curvature, we will see this momentarily.
\paragraph{}
{\bf Instanton solutions:} In addition to the smooth Coulomb branch-like solutions stated in (\ref{Saddle:Coulomb}), at North pole $\rho=0$ and South pole $\rho=\pi$ of $S_{b^2}^4$, we can have singular gauge configuration which are precisely the self-dual Yang-Mills instanton and anti-self-dual instanton solutions, which gives the non-perturbative contributions to the partition function.  
Explicitly we notice that if we allow for $F_{mn} \neq 0$, we can still satisfy (\ref{BosVecDeform}) by having $F_{mn}\bsig^{mn} = 0$ at $\rho=0$, i. e. anti-self-dual Yang-Mills instanton solution at north pole or $F_{mn}\sig^{mn}=0$ at $\rho = \pi$  i.e. self-dual Yang-Mills instanton solutions at south pole.
Moreover as mentioned earlier around the North and South poles, the metric reduces to that of $\Omega$-background, 
one can readily compute the contributions from these additional saddle point solutions using the instanton partition functions in \cite{Nekrasov:2002qd}.
\paragraph{}
Turning on non-trivial real valued functional $H(\phi, q)$, in particular it contains 4D FI parameter $\zeta \in {\mathbb R}^+$ and for concreteness, 
we consider for the time being $H(\phi, q) =  H(q)$ i. e. independent from the vector multiplet scalar and take the following form: 
\be\label{Def:Hq}
H(q) = (q_{\hI A})^{\dag} q_{\hI A}-\zeta, 
\ee
where $\zeta = \vec{\zeta}\cdot\vec{h}$ sums over the Cartan generators of the gauge group.
We will also restore the complex mass parameters $(\mu,\bar{\mu})$ from now on.
In the presence of $H(q)$ we will continue to set $\phi=\bphi=-\frac{i}{2} a_0$
such that  $[\phi,\bar{\phi}]^2$ and other $(\phi-\bar{\phi})$ dependent terms in (\ref{IvecHiggs-Bosonic}) and (\ref{IHyp:Bosonic}) vanish identically,
this is also consistent with the condition:
\be\label{DmphiCond}
D_m (\phi+\bphi) = 0
\ee
provided the resultant non-trivial gauge field $A_m$ also take values in the Cartan of gauge group. 
For the remaining terms in $\cI_{\rm vec.}+\cI_{\rm Higgs}\mid_{\rm Bosonic}$ (\ref{IvecHiggs-Bosonic}) to also vanish globally however, 
along with the D-term constraint (\ref{Dterm-constraint}), we need to impose on the gauge fields:
\bea&&(F_{32}+F_{41}) = \cos\theta H(q), \quad (F_{31}+F_{42}) = \sin\theta H(q), \quad (F_{12}+F_{34}) =0,\label{SaddleEqns4}
\\
\label{SaddleEqns3}
&&(F_{32}-F_{41}) = \cos\theta H(q), \quad (F_{31}-F_{42}) = \sin\theta H(q), \quad (F_{12}-F_{34}) = 0.
\eea
We can now classify the solutions to (\ref{SaddlePtEqn: General}) based on the possible solutions for $\cI_{\rm hyp.}|_{\rm Bose}$ (\ref{IHyp:Bosonic}) to vanish.
\paragraph{}
{\bf Deformed Coulomb solutions:}
The simplest solution to $\cI_{\rm hyp.}=0$ is to simply to set $q_{\hI A} = 0$, for (\ref{SaddleEqns3}) and (\ref{SaddleEqns4}) to vanish we need to have
\footnote{Strictly speaking, as pointed out in \cite{Pan:2015hza}, in addition to set the imaginary part of $\phi$ to be constant, 
the real of $\phi$ also needs to acquire additional spatial dependence proportional to $\zeta$ for Bianchi identity to be satisfied. 
However the crucial point here is that when we compare with the contour integration from Coulomb branch localization in the next section, 
we need to take $\zeta \to \infty$ limit such that the contributions from this branch are suppressed, the essential physics does not change due to this correction.}
:
\bea\label{DefCol-solution}
&&F_{12} = F_{34} = F_{41}=F_{42} = 0,\quad F_{13} = \zeta\sin\theta, \quad F_{23} = \zeta\cos\theta,\\
&&{\rm or} ~ F = \zeta (\cos\theta E^1 +\sin\theta E^2)\wedge E^3.\label{ConstF}
\eea
It is also worth noticing that when restricting $F$ to $\theta=0$ or $\theta=\frac{\pi}{2}$, $E^3$ degenerates and (\ref{ConstF}) vanishes identically,
in other words it does not mix with the vortex flux restricted along $S_b^2$ or $S^2_{1/b}$.
Moreover $\zeta$ hence the field strength $F_{mn}$ is proportional to identity in the color space, this is consistent with (\ref{DmphiCond}).
\paragraph{}
{\bf Higgs branch-like and Vortex solutions:}
Next we consider the possibility $q_{\hI A}$ picks up a non-vanishing constant expectation value $\sim \sqrt{\zeta}$ for $H(q)$ to vanish,
first we note from (\ref{SaddleEqns3}) and (\ref{SaddleEqns4}) that $F_{mn}=0$, thus we can set $A_m=0$ via gauge transformation, 
and these two sets of equations are satisfied trivially.
Now to see this is also a possible solution for $\cI_{\rm hyp.}|_{\rm Bose}$ (\ref{IHyp:Bosonic}) to vanish, let us first take the simplified limit of $S^4$, i. e. $r= l = \tilde{l}$, 
the geometric factor $h$, the background auxiliary tensor field $T_{mn}$, $\bar{T}_{mn}$ and $SU(2)_R$ gauge field $(V_m)_B^{~A}$ all vanish identically. 
We see that $D_m q_{\hI A}$ as defined in (\ref{Def:Dq}) now vanishes for constant $q_{\hI A}$.
However the anti-symmetric auxiliary tensor fields $S_{mn}$ and $\bar{S}_{mn}$ tend to non-vanishing constants 
and they give a finite mass to $q_{\hI A}$ through ${\mathbb M}$ (\ref{MassMatrix1}) and $\cM_{AB}$ (\ref{MassMatrix2}).
For $\cM_{AB}$ dependent terms to vanish, we can further demand additional complex mass parameters to satisfy $\mu = \bar{\mu}=i \frac{m_0}{2} \in i{\mathbb R}$ 
and
\be\label{MassCond}
(\phi+\mu) =(\bphi+\bar{\mu}) = \frac{i}{2}(m_0-a_0). 
\ee
Finally to have vanishing ${\mathbb M}$, such that hypermultiplet can become massless we need to analytically continue the value of $a_0$ to complex plane such that:
\be\label{HiggsCond}
a_0 = m_0  \mp \frac{i}{2}  \left(\frac{1}{f_0}+\frac{1}{g_0}\right).
\ee
where $f_0=g_0 = r$ are both constants.
This analytic continuation will be justified when we evaluate the partition function on $\cZ_{S_{b^2}^4}$,  
the condition (\ref{HiggsCond}) labels the lowest of infinite set of simple poles in the contour integration, 
and final expression becomes holomorphic function of $m_0 \mp \frac{i}{r}$. 
The same conclusion can be reached from the equations (\ref{HyperSaddle1}) and (\ref{HyperSaddle2}).
In particular we notice either $-$ or $+$ choice in (\ref{HiggsCond}) corresponds respectively 
to the condition for either $q_{ \hI 1}$ or $q_{\hI 2}$ of $SU(2)_R$ doublet becomes massless and picks up non-vanishing constant expectation value, 
while the other component is frozen at zero. 
We refer to this class of smooth solutions as ``Higgs-like'' since $H(q)$ can vanish, 
precisely which component vanishes is determined by the sign of the FI-parameter $\zeta$.
\paragraph{}
In addition to these smooth solutions, near the north and south poles ($\rho=0$ and $\rho=\pi$) 
we can also have the non-trivial singular vortex-like solutions such that 
$A_n$ and the non-vanishing $q_{\hI A}$\footnote{That is, either $q_{1 A}$ or $q_{2A}$ is non-vanishing while the other is set to zero.} 
become dynamical fields while the remaining parameters satisfy (\ref{MassCond}) and (\ref{HiggsCond}).
Let us now focus near $\rho=0$, an analogous discussion can be given for near $\rho={\pi}$. 
As mentioned earlier that the geometry near $\rho=0$ reduces to 4d $\Omega$-background\footnote{For the moment we have round $S^4$ such that $\epsilon_1=\epsilon_2=\frac{1}{r}$.},
non-trivial gauge configurations need to satisfy the equations (\ref{SaddleEqns4}) 
and the non-vanishing $q_{\hI A}$ component needs to satisfy the equation:
\be\label{Dq=0} 
D_n q_{\hI A}=0
\ee 
for the combined deformation terms to vanish, i.e. (\ref{SaddlePtEqn: General}) is satisfied.
The combined set can be regarded as the generalization of the mixed instanton-vortex equations found in \cite{Hanany:2004ea} to the 4d $\Omega$-background for $\epsilon_1=\epsilon_2$
\footnote{In \cite{Pan:2015hza}, the authors discovered another class of saddle point solutions corresponding to non-perturbative topological solitons which they called 
``Seiberg-Witten monopoles''. However the fluctuation determinant around these saddle points seem to be identical to the ones for the instantons in the presence of vortices we considered here, i. e. given by direct substitution of poles of perturbative one-loop determinant into the non-perturbative Nekrasov instanton partition function, 
as we will demonstrate explicitly in next section.  In particular as reviewed in Section 5.3 of \cite{Marino:1996sd} that,  on a Kahler manifold as it is the case of $S_{b^2}^4$, non-abelian Seiberg-Witten monopole equations reduce to vortex equations which take similar form to the vortex-instanton equations (\ref{SaddleEqns4}) and (\ref{SaddleEqns3}), It would therefore be very interesting to understand the connection of these two different classes of topological solitons.}
.
\paragraph{}
For general $S_{b^2}^4$ i. e.  $l \neq \tilde{l}$, $T_{mn}$, $\bar{T}_{mn}$, $(V_m)_B^{~A}$ and $h$ are non-vanishing, 
$f$ and $g$ also become non-constant, it is easier to consider linear equations (\ref{HyperSaddle1}) and (\ref{HyperSaddle2}).
First we note that if one of the $SU(2)_R$ doublet, say $q_{2\hI}$ remains zero, while the other non-vanishing component $q_{1\hI}$ still satisfies (\ref{Dq=0}), 
it is sufficient to ensure $\cI_{\rm hyp.}\mid_{\rm Bose}$ vanishes.
To solve $D_n q_{1\hI} =0$, we can consider the ansatz $q_{1\hI}(y)=\exp(-i \int (V_n)_1^{~1} dy^n) {\bf q}_{1\hI}$, 
such that  ${\bf q}_{1\hI}$ satisfies the simplified equation: $\partial_n {\bf q}_{\hI 1}-i(A_n)_{\hI}^{~\hat{J}}{\bf q}_{\hat{J}1}=0$.
We see that while ${\bf q}_{\hI 1}$ can still take the non-vanishing constant value when $A_n=0$,  
$q_{1\hI}$ becomes spatial dependent due to non-constant factor $\exp(-i \int (V_n)_1^{~1} dy^n)$. 
However we would like to argue that when we approach $\rho=0$ (similarly for $\rho=\pi$), 
$q_{1\hI}$ can still approach to non-vanishing constant expectation values such that $H(q)$ vanishes.
This can be seen from the explicit $SU(2)_R$ background field strength near $\rho=0$ computed in \cite{Pestun:2014mja}, 
which implies relevant component becomes:
\be\label{V11}
(V_n)^{~1}_{1} dy^n = \frac{1}{2}\rho^2\left[ \left(1-\frac{r^2}{l^2}\right) \cos^2\theta d\varphi+  \left(1-\frac{r^2}{\tl^2}\right) \sin^2\theta d\chi\right]
\ee
and $\lim_{\rho\to 0} \exp(-i \int (V_n)_1^{~1} dy^n) \to 1$ smoothly. We conclude that we can have isolated ``Higgs vacua'' $H(q)=0$ at $\rho=0$ (and $\rho=\pi$).
\paragraph{}
More generally in the presence of $A_n$ satisfying (\ref{SaddleEqns3}) and (\ref{SaddleEqns4}), 
 $q_{1\hI }$ can again become dynamical, subjected to the boundary condition that it approaches to non-vanishing constant at $\rho=0$ (or $\rho=\pi$).  
Near $\rho=0$ (or $\rho=\pi$), the combined set of equations describes 
the generalization instanton-vortex (anti-instanton-vortex) mixture configurations in 4d $\Omega$ background with arbitrary $\epsilon_{1,2}$\footnote{See \cite{Bulycheva:2012ct} for a recent discussion on vortex equations in $\Omega$-background.}.
When we explicitly evaluate the contour integral for $\cZ_{S^4_{b^2}}$ in the next section,
the resultant residues can be identified with the partition functions of the corresponding 2d $\cN=(2,2)$ vortex world volume theory, defined on the two deformed two spheres $S_{b}^2$ and $S^2_{1/b}$. We can regard the answer as having two distinct sets of vortex configurations of different topological charges wrapping along $S_b^2$ and $S_{1/b}^2$. They intersect each other at their common north and south poles, so from the perspective of $S_{1/b}^2$, the vortex configuration wrapping on $S_b^2$ also appears as co-dimension two defects located its north and south poles\footnote{From this perspective, the boundary condition for $q_{1\hI}$ at $\rho=0$ for example, can be regarded as the $S_{b^2}^4$ analogue of the condition that scalar field approaches non-vanishing vev $\sqrt{\zeta}$ for $H(q)=0$ at spatial infinity from the center of vortices now located at $\rho=\pi$; vice versa for the boundary condition at $\rho=\pi$.}. 
More precisely as we will provide supporting evidence later, they combine with the four dimensional instantons to appear as the generalized non-perturbative ``world sheet instanton'' corrections in $\cZ_{S^2_{1/b}}$.


\section{Explicit evaluation of ${\mathcal Z}_{S_{b^2}^4}$ for $\cN=2$ SQCD}\label{Sec:4dEvaluation}
\paragraph{}
In this section, we perform explicit contour integration for the $S_{b^2}^4$ partition of  
$\cN=2$ SQCD with $U(N_{c})$ gauge group and $N_f$ fundamental hypermultiplets, 
focusing on the relevant singularities and compute their residues. 
We will observe that the resultant expression exhibits almost factorizable structures, 
different contributions can be identified with the different saddle points solutions discussed earlier 
and we will discuss how they can be interpreted as 2d effects in the vortex world volume.
\paragraph{}  
Let us first write down the general expression for $S_{b^2}^4$ partition function for $\cN=2$ SQCD following \cite{Hama:2012bg}:   
\begin{eqnarray}\label{Z: 4dSQCD}
{\cal Z}^{\rm SQCD}_{S_{b^{2}}^4}&=&\frac{1}{N_c !}\int \prod_{\ra=1}^{N_c} d \hat{a}_{\ra}~e^{-\frac{8\pi^2}{g^2_{\rm YM}}{\sum_{\ra=1}^{N_c}}\ha_\ra^2+16i\pi^{2}\hat{\zeta}{\sum_{\ra=1}^{N_c}}\ha_\ra}\frac{\prod_{\ra\neq\rb=1}^{N_c}\Upsilon(i(\ha_\ra-\ha_\rb))}{\prod_{\ra=1}^{N_c}\prod_{I=1}^{N_f}\Upsilon\left(i(\ha_\ra+\hmu_I)+\frac{Q}{2}\right)} |\cZ_{\rm inst}(\vec{\ha},\vec{\mu}, q)|^2\nn\\
\end{eqnarray}
This expression was obtained through Coulomb branch localization discussed earlier.
Let us comment on the various contributions in (\ref{Z: 4dSQCD}), which was computed in the absence of the ${\bf Q}$-exact Higgs branch deformation (\ref{HiggsDeform}).
The first set of contributions are the classical pieces coming from evaluating the supersymmetric Lagrangians (\ref{Lvec}), (\ref{LFI}) and (\ref{Lhyp}) at the smooth perturbative Coulomb branch saddle points (\ref{Saddle:Coulomb}). 
Here $U(N_c)$ gauge group admits non-vanishing FI parameter $\zeta$, and we defined the dimensionless normalized values:
\be\label{Def: normalized}
\ha_\ra = \sqrt{l\tl} a_\ra = \frac{a_{\ra}}{\sqrt{\eps_1\eps_2}},\quad \hat{\zeta} = \sqrt{l\tl}\zeta = \frac{\zeta}{\sqrt{\eps_1\eps_2}},  
\quad \hat{\mu}_I = \sqrt{l\tl}\mu_I = \frac{\mu_I}{\sqrt{\eps_1\eps_2}}
\ee
where $\{a_{\ra} \in {\mathbb R}\}$ are the vevs of $N_c$ scalars in the Cartan of $U(N_c)$ and $\{\mu_I\}$ are the $N_f$ complex mass parameters.
The second set of contributions come from the zero mode fluctuation determinants of vector and hyper- multiplets around the (\ref{Saddle:Coulomb}), 
these are encoded through the function $\Upsilon(x)$:
\be\label{Def:Upsilon}
\Upsilon(x)= \prod^{\infty}_{m,n=0}(x+mb+nb^{-1})(x-mb-nb^{-1}-Q),\quad Q = b+\frac{1}{b} = \frac{l}{r} +\frac{\tl}{r}.
\ee
where we have also related the dimensionless deformation parameters $b$ and $1/b$ with the lengths $(r, l, \tl)$.
The vector and hyper-multiplets now contribute respectively to the numerator and denominator, 
we therefore expect them to contribute simple zeros and simple poles in the contour integration.
For later purpose, let us also define the following function
\be
\label{Def:gammax}
\gamma(x) = \frac{\Gamma(x)}{\Gamma(1-x)},
\ee
where $\Gamma(x)$ is the usual Gamma function and is related to $\Upsilon(x)$ through the following useful identity:
\be\label{UpGa-Identity}
\frac{\Upsilon(x+mb+{n}/{b})}{\Upsilon(x)} = (-1)^{mn}\prod_{r=0}^{m-1}\frac{\gamma(b(x+r b))}{b^{2b(x+rb)-1}} 
\prod_{s=0}^{n-1}\frac{\gamma(b^{-1}(x+s b^{-1}))}{(b^{-1})^{2b^{-1}(x+sb^{-1})-1}}
\prod_{r=0}^{m-1}\prod_{s=0}^{n-1} [(x+r b + s b^{-1})^2].
\ee
where $m, n \in {\mathbb Z}^{+}$. This identity will be useful when comparing 4d and 2d partition functions.
\paragraph{}
Finally the third set of contributions arise from the zero mode fluctuations around the singular non-perturbative anti-instanton/instanton saddle points $F^{\pm} = 0$ localized at north and south poles of $S^4_{b^2}$ respectively\footnote{We choose this convention to be consistent with the positions of instanton/anti-instanton saddle points in Section \ref{Sec:HiggsLoc}.}, 
they consist of two copies of the instanton partition functions computed in $\Omega$-background \cite{Nekrasov:2002qd}, \cite{Nekrasov:2003rj}.
There are many mathematically equivalent ways to express the instanton partition functions following the useful results in \cite{Awata:2008ed}, \cite{Sulkowski:2009ne}, for our purpose we choose the following representations:
\bea
&&{\cal Z}_{\rm inst}(\vec{\ha}, \vec{\hmu}, q)=\sum_{\{\vec{Y}\}}q^{|\vec{Y}|} \prod_{\ra,\rb=1}^{N_c} Z^{\rm vec}_{\ra\rb}(\vec{\ha},\vec{Y})\prod_{I=1}^{N_f}\prod_{\ra=1}^{N_c} Z^{\rm hyp}_{\ra I}(\vec{\ha},\vec{\hmu},\vec{Y}), \label{Def:Zinst}  \\
&&Z^{\rm vec}_{\ra \rb}(\vec{\ha}, \vec{Y}) =\prod^{\infty}_{r,s=1}\frac{\Gamma\left(Y_{\ra r}-Y_{\rb s}+b^{2}(r-s-1)+ib(\hat{a}_{\ra}-\hat{a}_{\rb})\right)}{\Gamma\left(Y_{\ra r}-Y_{\rb s}+b^{2}(r-s)+ib(\hat{a}_{\ra}-\hat{a}_{\rb})\right)}\frac{\Gamma\left(b^{2}(r-s)+ib(\hat{a}_{\ra}-\hat{a}_{\rb})\right)}{\Gamma\left(b^{2}(r-s-1)+ib(\hat{a}_{\ra}-\hat{a}_{\rb})\right)}, \label{Def:Zinst-vec}\nn\\
\\
&&Z^{\rm hyp}_{\ra I}(\vec{\ha},\vec{\hmu}, \vec{Y}) =\prod_{r=1}^{\infty}
\frac{\Gamma(b(i(\ha_\ra+\hmu_I)+\frac{Q}{2})+b^2(r-1)+Y_{\ra r})}{\Gamma(b(i(\ha_\ra+\hmu_I)+\frac{Q}{2})+b^2(r-1))}. \label{Def: Zinst-hyp}
\eea
Here $q=\exp(2\pi i\tau)$ and $\tau\equiv\frac{\theta}{2\pi}+\frac{4\pi i}{g^{2}_{\rm YM}}$ is the complex coupling. 
We use $\vec{Y}=(Y_{1},Y_{2},\ldots,Y_{N_{c}})$ to denote a set of $N_c$ Young diagrams and each Young diagram $Y_{\ra}$ is characterized by a set of integers $(Y_{\ra1}\geq Y_{\ra2}\geq\cdots\geq 0)$, 
where $Y_{\ra r}$ is the length of the r-th column
\footnote{Alternatively we can consider the transposed Young diagrams $\vec{Y}^{\vee}$, characterized by a set of integers $(Y^{\vee}_{\ra1} \geq Y^{\vee}_{\ra2} \geq \cdots \geq 0)$ where $Y_{\ra s}^{\vee}$ is the length of s-th row, we also need to exchange $b$ and $b^{-1}$ in the partition function.}. Moreover $|\vec{Y}|=\sum_{\ra,r}Y_{\ra r}$ denotes the total number of instantons and when $\vec{Y}$ is empty $\cZ_{\rm inst}(\vec{\ha},\vec{\hmu}, q)$ reduces to 1.
If we regard (\ref{Def:Zinst}) arise from the zero modes around the instantons localized at South pole $\rho=\pi$, then the contributions from those around the anti-instantons localized at North pole $\rho=0$ can be obtained from taking $q \to \bar{q} = \exp(-2i\pi\bar{\tau})$ in $\cZ_{\rm inst}(\vec{\ha},\vec{\hmu}, q)$. 
They combine to form the non-perturbative contributions $|\cZ_{\rm inst}(\vec{\ha}, \vec{\hmu}, q)|^2$ in (\ref{Z: 4dSQCD}).
\paragraph{}
To evaluate $\cZ^{\rm SQCD}_{S^4_{b^2}}$ explicitly, we need to promote $a_{\ra}$ into complex variables, the integration contour depends crucially on the sign and the magnitude of the FI parameter $\zeta$. Let us take $\hat{\zeta} \in {\mathbb R}^+$ for definiteness, this implies that we should choose the integration contour to be a semi-circle in the upper half plane for possible convergence. If we express $a_{\ra} = |a_{\ra}| e^{i\theta_\ra}$, the classical exponential factor can only ensure the contributions from the circular arc of radius $|a_\ra|$ to vanish exponentially, if we also have $\hat{\zeta} \ge \frac{|a_\ra| \cos2\theta_\ra}{2 g_{\rm YM}^2 \cos\theta_\ra}$. 
We therefore conclude that to take $|a_{\ra}| \to \infty$ limit such that the integration contour extends along the entire real line and encloses all the upper-half plane, we need to simultaneously take $\hat{\zeta} \to \infty$. 
Next let us consider the possible singularities in the upper half plane and their physical interpretation. 
They come from the simple poles in hypermultiplet perturbative zero mode fluctuation determinant, 
which we can read off their position from the definition of (\ref{Def:Upsilon}):
\be\label{Def: PoleConds}
\ha_\ra = -\left(\hmu_{l_\ra} -i\frac{Q}{2}\right)+i\left(m_{\ra} b+\frac{n_{\ra}}{b}\right), \quad l_\ra \in \{I \}, ~~ m_{\ra}, n_{\ra} = 0, 1, 2, 3, \dots
\ee
where the index $l_\ra$ implies that we pick $N_c$ out of $N_f$ possible complex mass parameters $\{\hmu_I \}$ which we assume to be all distinct.
To understand the physical meaning of these simple poles, we can use the related discussion in \cite{Gaiotto:2012xa} for the superconformal index to our $S_{b^2}^4$ partition function (see also \cite{Chen:2011sj} for earlier discussion). 
The basic idea is similar to the singularities in the Wilsonian effective action which can be attributed to certain fields accidentally become light and can condense to acquire expectation values. 
As the result, the local and global symmetries these fields transform under become partially if not completely broken. 
Moreover if we allow for the expectation of a given field to be spatial dependent, space-time rotational/translation symmetries hence supersymmetries can also be partially broken. We can regard this phenomenon as the insertion of extended topological BPS defects into our theory and the corresponding spatial dependent expectation values vanish precisely at their positions.
\paragraph{}
We can now associate the simple pole conditions (\ref{Def: PoleConds}) with the various saddle point conditions found earlier.
For $m_\ra, n_\ra = 0$, this corresponds to the locus in the Coulomb branch moduli space where $N_c$ fundamental scalars can become light and acquire non-vanishing expectation values, up to normalization this is precisely the saddle point condition found in (\ref{HiggsCond}), i. e. isolated Higgs vacua.
The shift by $\frac{i}{2}Q = \frac{i}{2}(\frac{l}{r}+\frac{\tilde{l}}{r})$ indicates the theory is defined on curved manifold $S_{b^2}^4$.
For $m_{\ra}, n_{\ra} \neq 0$, they correspond to inserting two sets of co-dimension two surface defects of topological winding number $m_\ra$ and $n_\ra$ respectively in the two orthogonal planes, which are the infra-red limit of the dynamical vortices (see e. g. \cite{Tong:2008qd} for an introduction.). 
The positive integers $m_\ra$ and $n_\ra$ also label the quantized angular momentum carried by the two sets of surface defects in their two transverse dimensions, 
these non-trivial topological configurations correspond to the first two equations of the saddle point equations (\ref{SaddleEqns3}) and (\ref{SaddleEqns4}).
The expectation values of the light scalar fields arising here parameterize the positions of the surface defects in space-time and internal spaces, 
in other words they become the moduli space coordinates of these surface defects, which will be promoted to 2d fields later. 
\paragraph{}
We can now readily evaluate the residues of $\cZ_{S_{b^2}^4}^{\rm SQCD}$ (\ref{Z: 4dSQCD}) at the the simple poles given in (\ref{Def: PoleConds}), and they are expressed in the convenient form to compare with the two dimensional vortex partition functions on $S_b^2$ and $S_{1/b}^2$ in the next section:
\be\label{Z: 4dSQCD-res}
\cZ_{S_{b^2}^4}^{\rm SQCD} = \left(\sum_{m_\ra =0}^{\infty} \sum_{n_\ra=0}^{\infty}
{\mathbb Z}^{\rm class.}_{\{m_\ra; b\}}  {\mathbb Z}^{\rm class.}_{\{n_\ra; b^{-1}\}}
{\mathbb Z}_{\{m_\ra; b\}}^{\rm 1~loop} {\mathbb Z}_{\{n_\ra; b^{-1}\}}^{\rm 1~loop} |\bZ_{\{m_\ra, n_\ra \}}^{\rm inst.}|^2  \bZ_{\{m_\ra, n_\ra\}}^{\rm cross} \right)\times\bZ^{\rm 4d~free}.
\ee
Let us discuss the various contributions in turns, we have separated them into explicit $(m_\ra, n_\ra)$ dependent and independent parts. 
The first pair of factors in the summation are:
\be\label{Def: Zclass}
\bZ_{\{m_\ra; b\}}^{\rm class.} = \exp\left[\frac{16\pi^2}{g_{\rm YM}^2}\sum_{\ra=1}^{N_c}\left( i m_\ra \left(b\hmu_{\rla}-ib\frac{Q}{2}\right)+\frac{m_\ra^2 b^2}{2}\right)\right],
\ee
and $\bZ_{\{n_\ra, \frac{1}{b}\}}^{\rm class.}$ can be obtained from above by exchanging $m_\ra \leftrightarrow n_\ra$ and $b \leftrightarrow \frac{1}{b}$.
These come from part of the classical supersymmetric action in (\ref{Z: 4dSQCD}) 
\footnote{We also divided final expression by phase factor $\lim_{\hat{\zeta}\to\infty}\exp[16i\pi^2\hat{\zeta}\sum_{\ra=1}^{N_c} (-\hmu_\ra+i\frac{Q}{2}+i(m_\ra b+n_\ra/b))]$.}. 
The second pairs of factors in the summation $\bZ_{\{m_\ra, b\}}^{\rm 1~loop}$ are:
\be\label{Def:Z1loop}
\bZ_{\{m_\ra, b\}}^{\rm 1~loop} =\Omega_{\{m_\ra; b\}}
\prod_{\ra=1}^{N_c} \prod_{r=0}^{m_\ra-1}
\frac{\prod_{\rb=1}^{N_c}\gamma( ib(\hmu_{\rla}-\hmu_{\rlb}) +(r-m_\rb)b^2)}{\prod_{j\neq \rla}\gamma(1+(r+1)b^2+ib(\hmu_\rla-\hmu_{j}))},
\ee
\be
\Omega_{\{m_\ra; b\}} = \prod_{\rb=1}^{N_c}\left[ \prod_{\ra=1}^{N_c} \prod_{r=0}^{m_\ra-1} b^{1-2((r-m_\rb)b^2-ib(\hmu_\rlb-\hmu_\rla))}\times \prod_{j \neq \rlb}\prod_{r=0}^{m_\rb-1} b^{1-2((r-m_\rb)b^2-ib(\hmu_\rlb-\hmu_j))} \right]
\ee
and we can again obtain $\bZ_{\{n_\ra, \frac{1}{b}\}}^{\rm 1~loop}$ by exchanging $m_\ra \leftrightarrow n_\ra, b\leftrightarrow 1/b$.
They arise from part of the perturbative zero mode fluctuation determinants in (\ref{Z: 4dSQCD-res}), after 
readily applying the identity for $\Upsilon(x)$ and $\gamma(x)$ (\ref{UpGa-Identity}).
The remaining $(m_\ra, n_\ra)$ dependent parts from perturbative contributions: 
\bea
&&\prod_{\ra,\rb=1}^{N_c}\left[\frac{\prod_{r=0}^{m_\rb-1}\prod_{s=0}^{n_\rb-1}[(r-m_\ra)b+\frac{(s-n_\ra)}{b}+i(\hmu_\rlb-\hmu_\rla)]}{\prod_{r=0}^{m_\rb-1}\prod_{s=0}^{n_\ra-1}[(r-m_\ra)b+\frac{(s-n_\ra)}{b}+i(\hmu_\rlb-\hmu_\rla)]  
\prod_{r=0}^{m_\ra-1}\prod_{s=0}^{n_\rb-1}[(r-m_\ra)b+\frac{(s-n_\ra)}{b}+i(\hmu_\rlb-\hmu_\rla)]
}\right]^2\nb\\\nb\\
&&\times\pd^{N_{c}}_{\ra=1}\pd_{j\neq\rla}\pd^{m_{\ra}-1}_{r=0}\pd^{n_{\ra}-1}_{s=0}\left[i(\hmu_{j}-\hmu_{\rla})+(r-m_{\ra})b+\frac{(s-n_{\ra})}{b}\right]^{-2},
\label{1loopCancel}
\eea 
this can be shown to almost cancel completely with the $\vec{Y}$ independent parts of the non-perturbative zero mode fluctuation determinant in (\ref{Z: 4dSQCD}).
\paragraph{}
\begin{figure}
\centering
\includegraphics[width=65mm]{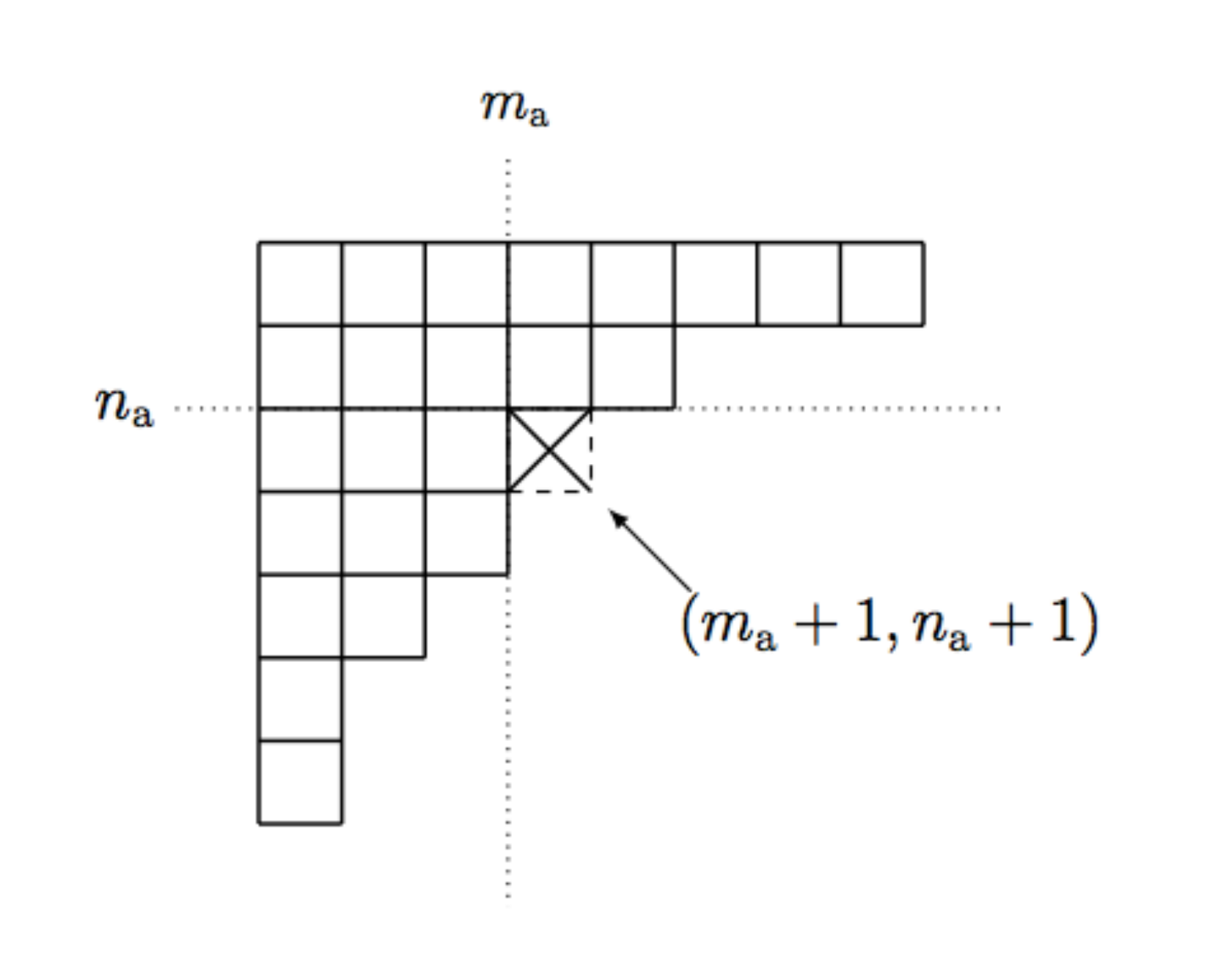}
\caption{The contributing Young Diagram $Y_\ra$ which avoids the box with coordinates $(m_\ra+1, n_\ra+1)$, 
the vertical or horizontal dotted lines indicate the division of $Y_\ra$ into $\oY_\ra$ and $Y_\ra/\oY_\ra$ and $\oY_\ra^\vee$ or $Y_\ra/\oY_\ra^\vee$ respectively.}
\label{Fig1}
\end{figure}
Before we present the explicit contributions from the non-perturbative parts (\ref{Def:Zinst}), let us discuss the contributing Young diagrams in the summation (\ref{Def:Zinst}).
We first notice that the infinite products of Gamma functions in (\ref{Def:Zinst}) 
need to be truncated to a finite products depending on set of pole numbers $\{m_\ra, n_\ra \}$.
This truncation arises from the hypermultiplet contribution (\ref{Def: Zinst-hyp}) which takes the following form upon the substitution of (\ref{Def: PoleConds}):
\be
Z_{\ra I}^{\rm hyp}(\vec{\ha},\vec{\hmu}, \vec{Y}) = \prod_{r=1}^{\infty} \prod_{s=1}^{Y_{\ra r}}[ib(\hmu_{I}-\hmu_\rla)+b^2(r-m_\ra-1)+(s-n_\ra-1)].
\ee
This implies that for given positive integers $(m_\ra, n_\ra)$, in order for the product above hence $\cZ_{\rm inst.}$ to be non-vanishing,
the Young diagram $Y_{\ra}$ cannot contain the box with coordinates $(m_\ra+1, n_\ra+1)$ as shown in Figure \ref{Fig1}, 
otherwise there will be zero in the product when $I=\rla$, $r=m_\ra+1$ and $s=n_\ra+1$. 
This also implies that when $m_\ra \neq 0, n_\ra=0$, to avoid the box $(m_\ra+1, 1)$, 
we need to truncate $Y_\ra$ at $m_a$ columns as indicated by vertical line in Figure \ref{Fig1} and we denote such a truncated Young diagram as $\oY_\ra$
\footnote{In our previous version, we mistakenly only included the truncated Young diagrams $\{\oY_\ra\}$ in our summation, we are grateful to Bruno Le Floch and the authors of \cite{Pan:2015hza} for pointing this out, and allowing to correct our subsequent related calculations.}.
Similarly for $m_\ra =0, n_\ra \neq0$, we need to truncate $Y_\ra$ horizontally at $n_\ra$ rows which can be denoted as $\oY_\ra^\vee$.
In order to facilitate the comparison with the partition function of vortex configuration wrapping on $S_b^2$ later,
we would like to divide $Y_\ra$ into $\oY_\ra$ which only contains first $m_\ra$ columns, and the complement of $\oY_\ra$ in $Y_\ra/\oY_\ra$
\footnote{One can consider alternative division of $Y_\ra$ into $\oY_\ra^\vee$  containing first $n_\ra$ rows and its complement $Y_\ra/\oY_\ra^\vee$,
this is suitable for comparison with the partition function of vortex configuration wraps on $S_{1/b}^2$ instead. }.
\paragraph{}
Let us now express the non-perturbative instanton contributions based on such vertical division of contributing Young diagrams $\{\vec{Y}\}$,
here we have also combined in our expressions (\ref{1loopCancel}) which almost cancel off the $\vec{Y}$ independent part from (\ref{Def:Zinst}), 
we are left with the following:
\bea\label{Def:Zinst-res}
\bZ^{\rm inst.}_{\{m_\ra, n_\ra\}} &=& \sum_{\{\vec{Y}\}} \bZ^{\rm inst.}_{2d}\times\bZ^{\rm inst.}_{\rm extra}.
\eea
The first factor in (\ref{Def:Zinst-res}) is given by:
\bea\label{Zinst<m}
\bZ^{\rm inst.}_{2d}&=& q^{|\vec{\bY}_{\rm 2d}|}(-1)^{N_c|\vec{\bY}_{\rm 2d}|} \prod_{\ra=1}^{N_c}\prod_{r=0}^{m_\ra-1}
\frac{\prod_{j\neq \rla}(i(\hmu_j-\hmu_\rla)b-b^2(r+1))_{\bY_{\ra(m_\ra-r)}}}{\prod_{\rb=1}^{N_c} (1+ib(\hmu_\rlb-\hmu_\rla)+b^2(m_\rb-r))_{\bY_{\ra (m_\ra-r)}}}\nonumber\\
&\times&\prod_{\ra,\rb=1}^{N_c}\frac{\prod_{r=0}^{m_\ra-1} (1+ib(\hmu_\rlb-\hmu_\rla)+b^2(m_\rb-r)+\bY_{\ra(m_\ra-r)}-\bY_{\rb 1})_{\bY_{\rb 1}}}
{\prod_{r=0}^{m_\ra-1}\prod_{r'=0}^{m_\rb-1}(1+ib(\hmu_\rlb-\hmu_\rla)+b^2(r'-r)+\bY_{\ra(m_\ra-r)}-\bY_{\rb(m_\rb-r')})_{\bY_{\rb(m_\rb-r')}-\bY_{\rb(m_\rb-r'+1)}}}
\nn\\
&\times&\prod_{\ra,\rb=1}^{N_c}\prod_{r=0}^{m_\ra-1}\frac{(1+ib(\hmu_\rlb-\hmu_\rla)-rb^2+\bY_{\ra(m_\ra-r)}-\bY_{\rb(m_\rb+1)})^{-1}_{\bY_{\rb(m_{\rb}+1)}}}{\prod_{s'=0}^{n_\rb-1}(i(\hmu_\rlb-\hmu_\rla)-(r+1)b+s' b^{-1}+\bY_{\ra(m_\ra-r)}b^{-1})}
\eea
where$(x)_n = \frac{\Gamma(x+n)}{\Gamma(x)}$, and we have introduced $\bY_{\ra (m_\ra-r)} =Y_{\ra (m_\ra-r)} -n_\ra=\oY_{\ra (m_{\ra}-r)}-n_\ra$, $r=0,\dots, m_\ra-1$ 
and $|\vec{\bY}_{\rm 2d}| = \sum_{\ra=1}^{N_c}\sum_{r=0}^{m_\ra-1}\bY_{\ra (m_\ra-r)}$.
This term consists exclusively of the boxes from each truncated Young diagram $\oY_\ra$, 
as the subscript ``2d'' indicated that this term will be identified with the instantons in the 2d vortex world sheet in the $n_\ra=0$ limit, such that the Young diagrams $\{Y_\ra\}$ now reduce to the truncated ones $\{\oY_\ra\}$, and the last ratio in (\ref{Zinst<m}) now also reduces to 1 as $Y_{\ra r}=0, \forall r\ge m_\ra+1$.
\paragraph{}
The remaining factors in the (\ref{Def:Zinst}) can now be packaged into:
\bea\label{Zinstextra}
\bZ^{\rm inst.}_{\rm extra}&=&q^{|\vec{Y}/\vec{\oY}|}(-1)^{N_{c}|\vec{Y}/\vec{\oY}|}\pd^{N_{c}}_{\ra,\rb=1}\Bigg{\{}\pd^{Y^{\vee}_{\ra 1}-m_{\ra}}_{r=1}\pd^{Y^{\vee}_{\rb 1}-m_{\rb}}_{s=1}\fc{\left(ib(\hmu_{l_{\rb}}-\hmu_{l_{\ra}})+b^2(r-s-1)-(n_{\ra}-n_{\rb})\right)_{Y_{\ra (r+m_{\ra})}-Y_{\rb (s+m_{\rb})}}}{\left(ib(\hmu_{l_{\rb}}-\hmu_{l_{\ra}})+b^2(r-s)-(n_{\ra}-n_{\rb})\right)_{Y_{\ra (r+m_{\ra})}-Y_{\rb (s+m_{\rb})}}}\nb\\\nb\\
&\times&\pd^{Y^{\vee}_{\ra 1}-m_{\ra}}_{r=1}\pd^{m_{\rb}}_{s=1}\Big{[}\fc{\left(ib(\hmu_{l_{\rb}}-\hmu_{l_{\ra}})+b^2(r-s-1)+m_{\rb}b^2-(n_{\ra}-n_{\rb})\right)_{Y_{\ra (r+m_{\ra})}-Y_{\rb s}}}{\left(ib(\hmu_{l_{\rb}}-\hmu_{l_{\ra}})+b^2(r-s)+m_{\rb}b^2-(n_{\ra}-n_{\rb})\right)_{Y_{\ra (r+m_{\ra})}-Y_{\rb s}}}\nb\\\nb\\
&\times&\fc{\left(ib(\hmu_{l_{\rb}}-\hmu_{l_{\ra}})+b^2(r-s-1)+m_{\rb}b^2-(n_{\ra}-n_{\rb})-Y_{\rb s}\right)_{Y_{\rb s}}}{\left(ib(\hmu_{l_{\rb}}-\hmu_{l_{\ra}})+b^2(r-s)+m_{\rb}b^2-(n_{\ra}-n_{\rb})-Y_{\rb s}\right)_{Y_{\rb s}}}\Big{]}\nb\\\nb\\
&\times&\pd^{m_{\ra}}_{r=1}\pd^{Y^{\vee}_{\rb 1}-m_{\rb}}_{s=1}\Big{[}\fc{\left(ib(\hmu_{l_{\rb}}-\hmu_{l_{\ra}})+b^2(r-s-1)-m_{\ra}b^2-(n_{\ra}-n_{\rb})\right)_{Y_{\ra r}-Y_{\rb (s+m_{\rb})}}}{\left(ib(\hmu_{l_{\rb}}-\hmu_{l_{\ra}})+b^2(r-s)-m_{\ra}b^2-(n_{\ra}-n_{\rb})\right)_{Y_{\ra r}-Y_{\rb (s+m_{\rb})}}}\nb\\\nb\\
&\times&\fc{\left(ib(\hmu_{l_{\rb}}-\hmu_{l_{\ra}})+b^2(r-s)-m_{\ra}b^2-(n_{\ra}-n_{\rb})\right)_{Y_{\ra r}}}{\left(ib(\hmu_{l_{\rb}}-\hmu_{l_{\ra}})+b^2(r-s-1)-m_{\ra}b^2-(n_{\ra}-n_{\rb})\right)_{Y_{\ra r}}}\Big{]}\nb\\\nb\\
&\times&\pd^{Y^{\vee}_{\ra 1}}_{r=m_{\ra}+1}\fc{1}{\left(ib(\hmu_{l_{\rb}}-\hmu_{l_{\ra}})+b^2(r-Y^{\vee}_{\rb 1}-1)-b^2(m_{\ra}-m_{\rb})-(n_{\ra}-n_{\rb})\right)_{Y_{\ra r}}}\nb\\\nb\\
&\times&\pd^{Y^{\vee}_{\rb 1}}_{s=m_{\rb}+1}\fc{1}{\left(ib(\hmu_{l_{\rb}}-\hmu_{l_{\ra}})+b^2(Y^{\vee}_{\ra 1}-s)-b^2(m_{\ra}-m_{\rb})-(n_{\ra}-n_{\rb})-Y_{\rb s}\right)_{Y_{\rb s}}}\Bigg{\}}\nb\\\nb\\
&\times&\pd^{N_{c}}_{\ra=1}\pd^{N_{f}}_{I=1}\pd^{Y^{\vee}_{\ra 1}}_{r=m_{\ra}+1}\left(ib(\hmu_{I}-\hmu_{l_{\ra}})+b^2(r-m_{\ra}-1)-n_{\ra}\right)_{Y_{\ra r}}
\eea
which consists of the contributions from entire Young diagram $Y_\ra$ and one can see this long expression become to 1 when $n_{\ra}=0$ such that $\vec{Y}=\vec{\oY}$ and $Y^{\vee}_{\ra 1}=m_{\ra}$. 
Also, to verify the validity of this long expression, 
we perform a simple test for the $U(1)$ gauge group case considered in \cite{Pan:2015hza} in Appendix \ref{U(1)}.
\paragraph{}
Finally, the remaining $(m_\ra, n_\ra)$ dependent terms from (\ref{Z: 4dSQCD}) which are independent of $\{\hmu_I\}$
can be packaged into:
\be\label{Z:cross}
\bZ^{\rm cross}_{\{m_\ra, n_\ra\}} = (-1)^{(2N_c-N_f)\sum_{\ra=1}^{N_c}m_\ra n_\ra}(b^2)^{\sum_{\ra=1}^{N_c}m_\ra\sum_{\rb=1}^{N_c}n_\rb+(N_f-2N_c)\sum_{\ra=1}^{N_c}m_\ra n_\ra},
\ee
these $(m_\ra, n_\ra)$-symmetric terms consist of classical and cannot be split into purely $m_\ra$ or $n_\ra$ dependent terms, 
moreover we notice that $\bZ^{\rm cross}_{\{m_\ra, 0\}} = \bZ^{\rm cross}_{\{0, n_\ra\}} =1$.
The last remaining factors consists of purely $(m_\ra, n_\ra)$ independent contributions:
\be\label{Z:4dfree}
\bZ^{\rm 4d~free} = \frac{ \exp\left[{-\frac{8\pi^2}{g_{\rm YM}^2}\sum_{\ra=1}^{N_c}\left(i\frac{Q}{2}-\hmu_\ra \right)^2}\right] }{\prod_{\ra=1}^{N_c}\prod_{j\neq l_\ra}\Upsilon\left(\frac{Q}{2}-i((\hmu_{\rla}-i\frac{Q}{2})-\hmu_j)\right)}.
\ee
We can identify the denominator here as the zero mode fluctuation determinant around the vanishing vevs of the free $N_c\times (N_f-N_c)$ 4d hypermultiplets.
We have written it in such a way to highlight the shift of $\hmu_{\rla} \to \hmu_{\rla}-i\frac{Q}{2}$ due to the Higgsing of 4d $U(N_c)\times SU(N_f)$ gauge and global symmetry group into residual $S[U(N_c)\times U(N_f-N_c)]$ residual global symmetry group\footnote{More precisely, the residual $S[U(N_c)\times U(N_f-N_c)]$ is further broken by the complex mass parameters into $U(1)^{N_f-1}$.} (See Figure \ref{Fig2}). 
We have also included in the numerator the classical action evaluated at the simple pole corresponding to the $N_c$ isolated Higgs vacua $\ha_\ra = -\hmu_\ra+i\frac{Q}{2}$. 
Finally, the complex conjugation for $\bZ^{\rm inst.}_{\{m_\ra, n_\ra\}}$ in (\ref{Z: 4dSQCD-res}) is again done simply by $q \to \bar{q}$.
\begin{figure}
\centering
\includegraphics[width=125mm]{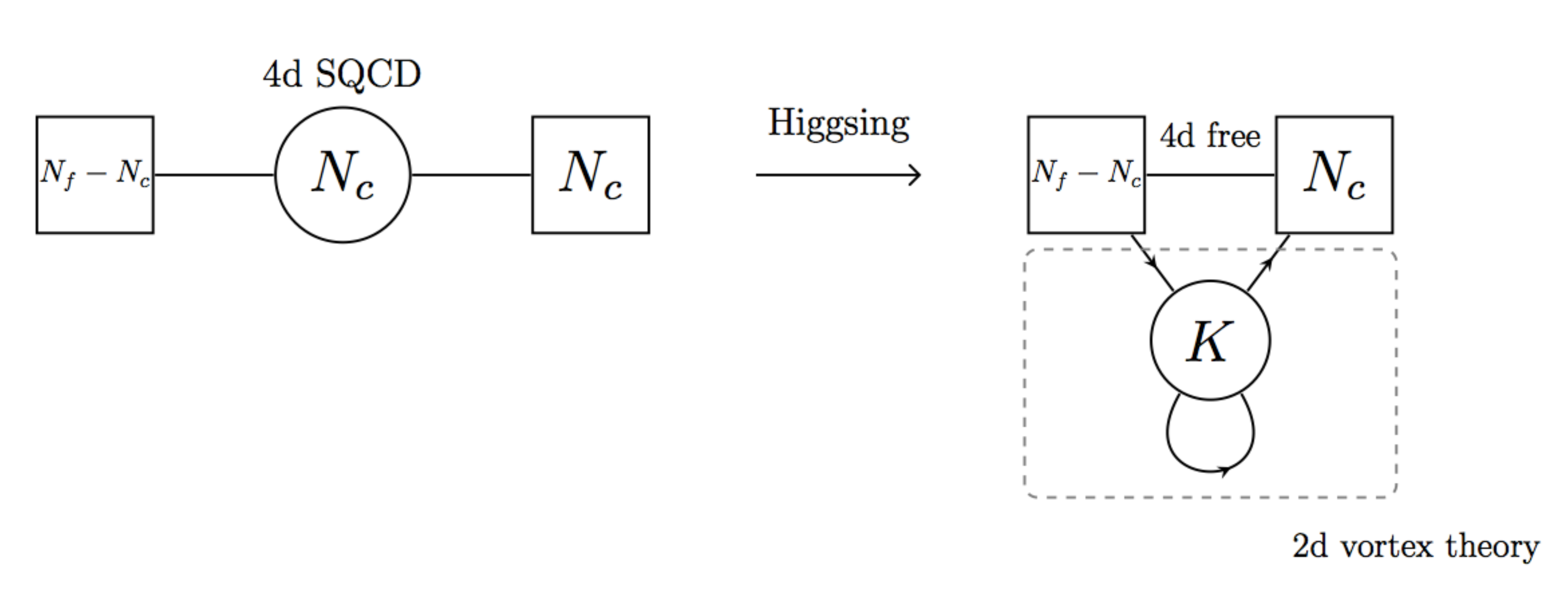}
\caption{Higgsing of 4d $\cN=2$ SQCD into 2d $\cN=(2,2)$ SQCDA coupled to free 4d hypermultiplets.}
\label{Fig2}
\end{figure}
\paragraph{}
Summarizing, the first four terms of $(m_\ra, n_\ra)$ dependent parts in (\ref{Z: 4dSQCD-res}) readily factorize into purely $(m_\ra, b)$ and $(n_\ra, b^{-1})$ parts consisting of $\bZ_{\{m_\ra, b\}}^{\rm class} \bZ_{\{m_\ra, b\}}^{\rm 1~loop}$ and 
$\bZ_{\{n_\ra, b^{-1}\}}^{\rm class} \bZ_{\{n_\ra, b^{-1}\}}^{\rm 1~loop}$,
this factorization is similar to what was observed in the context of 4d superconformal indices \cite{Peelaers:2014ima, Yoshida:2014qwa, Chen:2014rca}, 
in the next section we will identify them with the classical and one-loop contributions to the partition functions of 2d $\cN=(2,2)$ vortex theory on $S_b^2$ and $S_{1/b}^2$ respectively, coupling to free $N_c\times (N_f-N_c)$ 4d hypermultiplets.
Interestingly, in addition to the constant factor $\bZ^{\rm cross}_{\{m_\ra,n_\ra\}}$, 
we also have the non-trivial non-factorizable term $|\bZ_{\{m_\ra, n_\ra\}}^{\rm inst.}|^2$, 
which will be matched precisely with the 2d world sheet instanton/anti-instanton contributions in the $n_\ra= 0$ (or equivalently $m_\ra=0$) limit, 
we will also comment on the general $m_\ra, n_\ra \neq 0$ case in the next section. 
We should note however all these classical, perturbative and non-perturbative contributions we found in (\ref{Z: 4dSQCD-res}) are in accordance with the various saddle point solutions we found in the previous section using Higgs branch localization on $S_{b^2}^4$.

\section{Comparison with $S_b^2$ partition function of $\cN=(2,2)$ vortex theory}\label{Sec:2dEvaluation}
\paragraph{}
The relevant $\cN=(2,2)$ two dimensional vortex theory on ${\mathbb R}^2 \subset {\mathbb R}^4$ was derived explicitly in \cite{Hanany:2004ea} using D-brane constructions, for total topological charge $K$ configuration, the vortex theory has $U(K)$ gauge group whose gauge field is contained in the vector multiplet, its matter content consists of $N_c$ fundamental chiral multiplets with twisted masses $M_\ra$; $N_f-N_c$ anti-fundamental chiral multiplets with twisted masses $M_{\bj}$
and also one adjoint chiral multiplet which is denoted as $\bX$ and charged under the $U(1)$ rotation symmetry in the transverse ${\mathbb R}^2$.
We refer to this theory as ``2d SQCDA''.
\paragraph{}
Now putting 2d SQCDA on deformed two sphere $S_b^2$,  the twisted mass and a $U(1)_R$ charge $\bq$ can now combine into a complex twisted mass \cite{Doroud:2012xw,Benini:2012ui}, it is useful to define dimensionless quantity \cite{Gomis:2014eya}:
\be\label{TWmass}
\rM = l M + \frac{\bq}{2}
\ee
the partition function is holomorphic in such a combination. Moreover as we embed $S_b^2$ into $S_{b^2}^4$, 
the finite curvature now induces a dimensionless complex twisted mass $\rM_\bX$ for the adjoint chiral $\bX$. 
Let us now write down the partition function of SQCDA on $S^{2}_{b}$, following \cite{Gomis:2012wy}:
\bea\label{Z: 2dSQCDA}
&&{\cal Z}^{\rm SQCDA}_{S_b^2}=\frac{1}{K!}\sum_{\vec{\bB}\in {\mathbb Z}^{K}}\int_{{\mathbb R}^{K}}\left[\prod^{K}_{\rr=1}\frac{d\sigma_{\rr}}{2\pi}e
^{-4\pi i \br\sigma_{\rr}+i\uptheta \bB_{\rr}}\right]\left[\prod^{K}_{\rr<\rs}\left(\frac{(\bB_{\rr}-\bB_{\rs})^2}{4}+(\sigma_{\rr}-\sigma_{\rs})^2\right)\right]\times\nonumber\\\nonumber
&&\prod^{K}_{\rr=1}\left[\prod^{N_c}_{\ra=1}\frac{\Gamma(-i\sigma_{\rr}-i\rM_{\ra}-\frac{\bB_{\rr}}{2})}{\Gamma(1+i\sigma_{\rr}+i\rM_{\ra}-\frac{\bB_{\rr}}{2})}\prod^{N_f-N_c}_{\bj=1}\frac{\Gamma(i\sigma_{\rr}-i{\rM}_{\bj}+\frac{\bB_{\rr}}{2})}{\Gamma(1-i\sigma_{\rr}+i{\rM}_{\bj}+\frac{\bB_{\rr}}{2})}\right]
\prod^{K}_{\rr,\rs=1}\frac{\Gamma(-i\sigma_{\rr}+i\sigma_{\rs}-i\rM_{\bX}-\frac{\bB_{\rr}-\bB_{\rs}}{2})}{\Gamma(1+i\sigma_{\rr}-i\sigma_{\rs}+i\rM_{\bX}-\frac{\bB_{\rr}-\bB_{\rs}}{2})}.
\eea
Here the $\sigma_{\rr}$ is the scalar component of the vector multiplet which takes value in the Cartan sub-algebra of the gauge group; 
$\br$ is the 2$d$ FI parameter admitted by $U(K)$ gauge group, it forms complex combination $i\br+\frac{\uptheta}{2\pi}$ with the 2d theta angle $\uptheta$;
finally $\bB_{\rr} \in {\mathbb Z}$ is the quantized magnetic charge on $S_b^2$. 
In the absence of superpotential as it is the case here, the value of $U(1)_R$ charge is unfixed, 
however interesting polynomial type of superpotential for chiral multiplets can also be added to this theory \cite{Gomis:2014eya}, 
this fixes the $U(1)_R$ charges to definite values.
\paragraph{}
For the special superconformal case $N_f -N_c = N_c $, the $S_b^2$ partition function (\ref{Z: 2dSQCDA}) has been computed explicitly in \cite{Gomis:2014eya},
the generalization to arbitrary $N_f-N_c$ is straightforward, and it boils down to picking up the residues at the following values:
\be\label{2d-PoleConds}
i\sigma_{\ra \bmu } \pm \frac{\bB_{\ra\bmu}}{2}= - i\rM_{\ra}-i\bmu\rM_{\bX}+ k_{\ra\bmu}^{\pm}
\quad  1\le\ra\le N_c, ~~0\le\bmu< \hm_\ra , 
~~ k_{\ra\bmu}^{\pm}\ge 0.
\ee
Here the positive integers $\{\hm_\ra\}$ need to satisfy the constraint $\sum_{\rs=1}^{N_c} \hm_\ra = K$, in other words we have partitioned $K$ scalars into $N_c$
distinct sets,
and residues are only non-vanishing provided two sets of independent positive integers $k_{\ra\bmu}^{\pm}$ satisfy $k_{\ra\bmu}^\pm \ge k_{\ra(\bmu+1)}^\pm$ for all $(\ra, \bmu)$ and $\pm$ signs. 
Importantly, as noticed in \cite{Gomis:2014eya} that the integrand only depends on either combination $i\sigma_{\ra\bmu}+\frac{\bB_{\ra\bmu}}{2}$ or $i\sigma_{\ra\bmu}-\frac{\bB_{\ra\bmu}}{2}$, we expect the $k^{+}_{\ra\bmu}$ and $k^{-}_{\ra\bmu}$ dependences in the resultant expression from (\ref{Z: 2dSQCDA}) to decouple into two separate summations.
Let us now summarize the answers from explicit contour integrations as following:
\be
{\cal Z}^{\rm SQCDA}_{S_b^2}=\sum_{\{\hm_{\ra}\}}{\bfZ}^{\rm class.}\bfZ_{\{\hm_\ra\}}^{\rm 1~loop}|\bfZ_{\{\hm_\ra\}}^{\rm vort.}({\rm z})|^2.
\ee
Here the classical part ${\bfZ}^{\rm class}$ can be written in terms of ${\rm z} = e^{-2\pi\br+i\uptheta}$ as:
\be\label{Z:2dClass}
{\bfZ}^{\rm class}=({\rm z\bar{z}})^{-\sum_{\ra=1}^{N_c}\sum_{\bmu=0}^{\hm_\ra-1}(i\rM_{\ra}+i\bmu \rM_{\bX})}
=\exp\left[4\pi i\br\sum_{\ra=1}^{N_c}\left(\hm_\ra\left(\rM_\ra-\frac{\rM_{\bX}}{2}\right)+\frac{\hm_\ra^2}{2}\rM_\bX\right)\right],
\ee
whereas one-loop perturbative contribution is given by
\be\label{Z:2d1loop}
{\bfZ}_{\{\hm_\ra\}}^{\rm 1~loop}=\frac{\prod_{\rb=1}^{N_c}\prod_{\ra=1}^{N_c}\prod_{\bmu=0}^{\hm_\ra-1}\gamma(i(\rM_{\ra}-\rM_\rb)+i(\bmu-\hm_\rb) \rM_{\bX})}{\prod_{\bj=1}^{N_f-N_c}\prod_{\ra=1}^{N_c}\prod_{\bmu=0}^{\hm_\ra-1}\gamma(1+i{\rM}_{\ra}+i\rM_{\bj}+i\bmu \rM_{\bX})},
\ee
finally the non-perturbative world sheet instanton partition function is
\bea\label{Z:2dvortex}
{\bfZ}_{\{\hm_\ra\}}^{\rm vort.}({\rm z})&=&\sum_{k:(\ra, \bmu) \to \mathbb{Z}\geq 0}\left[(-1)^{N_{c}+K-1}{\rm z}\right]^{\sum_{\ra=1}^{N_c}\sum_{\bmu=0}^{\hm_\ra-1}k_{\ra\bmu}}\prod^{N_{c}}_{\ra=1}\prod_{\bmu=0}^{\hm_\ra-1}\frac{ \prod^{N_f-N_c}_{\bj=1}  (-i{\rM}_{\bj}-i\rM_{\ra}-i\bmu \rM_{\bX})_{k_{\ra\bmu}}}
{\prod_{\rb=1}^{N_c}(1+i\rM_{\rb}-i\rM_{\ra}+(\hm_{\rb}-\bmu)i\rM_{\bX})_{k_{\ra\bmu}}}\nn\\
&\times&
\prod^{N_{c}}_{\ra=1}\prod_{\bmu=0}^{\hm_\ra-1}
\frac{\prod^{N_{c}}_{\rb=1}(1+i\rM_{\rb}-i\rM_{\ra}+(\hm_{\rb}-\bmu)i\rM_{\bX}+k_{\ra\bmu}-k_{\rb(\hm_{\rb}-1)})_{k_{\rb(\hm_{\rb}-1)}}}
{\prod_{\rb=1}^{N_c}\prod_{\bnu=0}^{\hm_\rb-1}(1+i\rM_{\rb}-i\rM_{\ra}+(\bnu-\bmu)i\rM_{\bX}+k_{\ra\bmu}-k_{\rb\bnu})_{k_{\rb\bnu}-k_{\rb(\bnu-1)}}},
\eea
where we have also set $k_{\rb, -1} = 0$.
Notice that we have omitted the superscript ``$\pm$'' on $k_{\ra\bmu}^{\pm}$ in above, 
it should be understood however that in (\ref{Z:2dvortex}) if we set $k_{\ra\bmu} \equiv k_{\ra\bmu}^{+}$, 
for its complex conjugate $\bZ_{\{\hm_\ra\}}^{\rm vort.}({\rm \bar{z}})$ we should set $k_{\ra\bmu} \equiv k_{\ra\bmu}^{-}$ along with ${\rm z} \to {\rm \bar{z}}$\footnote{While keeping the mass dependence expression the same.}.
We can naturally interpret $k_{\ra\bmu}^+$ and $k_{\ra\bmu}^-$ respectively as the world sheet instanton and anti-instanton number, localized at south and north poles of $S_b^2$.  
\paragraph{}
We are now ready to compare the residues obtained from $\cZ_{S_{b^2}^4}^{\rm SQCD}$ as given in (\ref{Z: 4dSQCD-res}) 
and the $S_b^2$ partition of the vortex partition function $\cZ_{S_b^2}^{\rm SQCDA}$. 
Let us begin by considering the $m_\ra \neq 0; n_\ra=0$ contributions in the summation, such that $\bZ^{\rm class}_{\{n_\ra, b^{-1}\}}=\bZ^{\rm 1~loop}_{\{n_\ra, b^{-1]}}=\bZ^{\rm cross}_{\{m_\ra,0\}} =1$, $\bY_{\ra r} = Y_{\ra r} = \oY_{\ra r}$,
and the last line of (\ref{Zinst<m}) also becomes identity. 
We now identify the 2d and 4d parameters as follows. 
From the classical parts (\ref{Def: Zclass}) and (\ref{Z:2dClass}), they become identical if we set:
\be\label{Matching1}
\frac{4\pi}{g_{\rm YM}^2} = \br, \quad m_\ra = \hm_\ra, \quad  b^2 = i \rM_{\bX},\quad ib\hmu_\rla +b^{2}+\frac{1}{2} = i\rM_\ra, \quad \ra =1, \dots, N_c
\ee
Next we see that modulo the factor $\Omega_{\{m_\ra; b\}}$ in (\ref{Def:Z1loop})\footnote{We expect this factor can be absorbed by renormalization of gauge coupling, but it would be nice to understand its precise origin.}, 
it matches with (\ref{Z:2d1loop}) if we further set:
\be\label{Matching2}
-ib\hmu_{j}-\frac{1}{2} = i\rM_{\bj}, \quad j\in \{I\}/\{\rla\}, \quad \bj = 1, \dots, N_f-N_c.
\ee
These state that we naturally identify the 4d vortex numbers $\{m_\ra\}$ with the integers $\{\hm_\ra\}$ partitioning total vortex charge $K$, 
and the dimensionless 4d complex masses $\{\hmu_\rla, \hmu_j\}$ and 2d twisted masses $\{\rM_\ra, \rM_\bj\}$.
It is also interesting to note that the identification of $b^2 = i\rM_{\bX}$ was also considered in \cite{Gomis:2014eya} when identifying the insertion of surface operators in 4d $\cN=2$ superconformal QCD with the degenerate vertex operator in the correlation function of the dual Toda field theory.
Finally with these parameter matching through the classical and perturbative contributions, the non-perturbative contributions (\ref{Zinst<m}) and (\ref{Z:2dvortex})
can also be identified if we set
\be\label{Matching3}
\oY_{\ra(m_\ra-r)} = k_{\ra r}, \quad, r=0, 1,\dots, m_\ra-1,\quad \theta = \uptheta + (K-1)\pi,
\ee
in other words we identify the 4d Yang-Mills instanton and 2d world sheet instanton numbers as claimed earlier.
We have demonstrated in $n_\ra =0$ limit,  the residue obtained from $\cZ_{S_{b^2}^4}^{\rm SQCD}$ 
indeed to the $S_b^2$ partition function for 2d vortex theory $\cZ_{S_b^2}^{\rm SQCDA}$ coupled free 4d hypermultiplets, 
exactly the same matching can be performed if we exchange $(m_\ra, b, \oY_\ra)$ with $(n_\ra, b^{-1}, \oY_\ra^{\vee})$, and accordingly in the parameter matchings (\ref{Matching1}), (\ref{Matching2}) and (\ref{Matching3}).
\paragraph{}
Let us next discuss when $m_\ra, n_\ra \neq 0$ in (\ref{Z: 4dSQCD-res}), first we note that both classical $\bZ_{\{m_\ra, b\}}^{\rm class}$,  $\bZ_{\{n_\ra, b^{-1}\}}^{\rm class}$ 
and perturbative one loop $\bZ_{\{m_\ra, b\}}^{\rm 1~loop}$ and  $\bZ_{\{n_\ra, b^{-1}\}}^{\rm 1~loop}$ again precisely match with their 2d counterparts given in (\ref{Z:2dClass}) and (\ref{Z:2d1loop}) after exchanging $b$ and $b^{-1}$ appropriately.
For the non-perturbative contribution $\bZ^{\rm inst.}_{\{m_\ra, n_\ra\}}$, we now need to consider the full Young diagrams $\{Y_\ra\}$.
We can however still match the first two lines in (\ref{Zinst<m}) with (\ref{Z:2dvortex}), if we instead identify $\bY_{\ra (m_\ra-r)} = \oY_{\ra (m_\ra-r)}-n_\ra$ with the world sheet instanton number $k_{\ra \bmu}$ or in other words we shift the world sheet instanton number as $k_{\ra \bmu} \to k_{\ra\bmu} - n_\ra$.
This is the main reason we divide $Y_\ra$ into $\oY_\ra$ and $Y_\ra/\oY_\ra$ when presenting the residue of instanton partition function.
Now from the perspective of $S_b^2$, the presence of the surface defects wrapping on the other deformed two sphere $S_{1/b}^2$ which intersects with $S_b^2$ at $\rho=0, \pi$, is encoded in the shifted world sheet instanton number $\bY_{\ra(m_\ra-r)}$. The resultant configuration can be naturally regarded as the bound state of 4d Yang-Mills instantons labeled by $\oY_{\ra}$ and the surface defects of charge $n_\ra$, both appearing as the co-dimension two defects on $S_b^2$. This is precisely the vortex-instanton configuration we found in Section \ref{Sec:HiggsLoc} \footnote{
An equivalent way to understand this shift is to realize that when we only have surface defects wrapping on $S_b^2$, 
the total world sheet instanton number is given by $\sum_{\ra=1}^{N_c}\sum_{\bmu=0}^{\hm_\ra-1} k_{\ra\bmu} = \frac{1}{2\pi}\int F_{z\bar{z}}$, where $F_{z\bar{z}}$ is the field strength along the two dimensional $z$-plane attached to the south pole $\rho=\pi$. If we also have surface defects wrapping on $S_{1/b}^2$,  
$F_{z\bar{z}}$ needs to carry the additional component $\Delta F_{z\bar{z}}$ such that when integrating over $z$-plane which is also transverse to $S_{1/b}^2$,  
the additional component yields $\frac{1}{2\pi}\int \Delta F_{z\bar{z}} = -\sum_{\ra=1}^{N_c} n_{\ra}$. }.
However there remain 4d Yang-Mills instantons associated with $Y_\ra/\oY_\ra$  which cannot be interpreted as world sheet instantons on $S_b^2$,
and they are responsible for the extra contributions in $\bZ^{\rm inst.}_{\rm extra}$. They should be regarded as the summation of world sheet instantons in the world volume theory of vortices wrapping on $S_{1/b}^2$, whose number is given by $\oY_{\ra}^\vee$, plus the contact interactions between $S_b^2$ and $S^2_{1/b}$ at $\rho=0, \pi$. It would be very interesting to make such correspondence precise, and it is currently under investigation \cite{WIP}. 
\paragraph{}
It is also interesting to recall from \cite{Hanany:2004ea} that these 2d world sheet instantons can be regarded as the vortex configurations in $\cN=(2,2)$ gauge theories.
In particular they were explicitly showed in  \cite{Benini:2012ui} and  \cite{Doroud:2012xw}  as the alternative saddle point solutions in the Higgs branch localization computation on $S^2$.
It would be interesting to first reproduce (\ref{Z:2dvortex}) by computing the equivariant volume of the vortex moduli space of $\cN=(2,2)$ SQCDA on $S_b^2$ in $n_\ra=0$ limit. The corresponding computations have been done for some closely related theories obtained from the superpotential deformations of $\cN=(4,4)$ theory, such as so-called $\cN=(2,2)^*$ theory where a Yukawa-type superpotential $\sim {\rm Tr}(\tilde{Q}\bX Q)$ \cite{Benini:2012ui}, \cite{Fujitsuka:2013fga} is added to SQCDA considered here\footnote{Another good example is the soft-breaking where a mass term ${\rm Tr}\bX^2$ is added to SQCDA \cite{Yoshida:2011au}.}. 
The additional superpotential simplifies the matrix model for the world sheet instantons, as the zero modes associated with the adjoint chiral $\bX$ get lifted, 
the discrete supersymmetric vacua of the matrix model hence the resultant partition function is only labeled by a single set of $N_c$ integers partitioning $K$, c. f. \cite{Benini:2012ui}, \cite{Fujitsuka:2013fga}, we expect that the dependence on two sets of integers $\{\hm_\ra\}$ and $\{k_{\ra \bmu}\}$ in (\ref{Z:2dvortex}) will be generated by these additional adjoint zero modes.  
When $n_\ra \neq 0$, such a computation will also shed lights on the role of the additional terms in last line of  (\ref{Zinst<m}) from the 2d perspective.
From the perspective of 4d $\cN=2$ gauge theory, this is equivalent to study the moduli space of Yang-Mills instantons localized on two intersecting stacks of surface defects which are IR limit of dynamical vortices of charges $\{m_\ra\}$ and $\{n_\ra\}$ respectively.

\section{Discussions}\label{Sec:Discussion}
\paragraph{}
We would like to end by discussing some interesting possible future directions.
\paragraph{}
While we worked out explicitly the new saddle point solutions 
arising from the ${\bf Q}$-exact Higgs branch deformation term $\cI_{\rm Higgs}$ on $S_{b^2}^4$ in section \ref{Sec:HiggsLoc}, 
the obvious omission in our current work is the explicit computation of the zero-mode fluctuation determinants 
around them. The answer should consist of two parts, the first one is the zero mode fluctuation around purely the vortex/surface defects background configurations, 
similar to the computations in  \cite{Fujitsuka:2013fga},  \cite{Doroud:2012xw}; 
the other part involves computing the volume of the moduli space of the Yang-Mills instanton attached to surface defects, 
i. e. the moduli space of the solutions to (\ref{SaddleEqns3}), (\ref{SaddleEqns4}) and (\ref{Dq=0}), this computation as far as we know, has not been done in the literature. The results should respectively match with $\bZ^{\rm 1~loop}_{\{m_\ra, b\}}$ and $|\cZ^{\rm inst.}_{\{m_\ra, n_\ra\}}|^2$ in (\ref{Def:Zinst-res}) obtained from the contour integration in section \ref{Sec:4dEvaluation}, which in turns serve as the predictions.
\paragraph{}
In this work we focused on $\cN=2$ $U(N_c)$ SQCD on $S_{b^2}^4$, however we believe the almost factorizable structure we observe in section \ref{Sec:HiggsLoc} applies to the theories with other gauge groups and matter contents admitting non-vanishing FI parameter and Higgs branch.
The residue here presumably can also be interpreted as the $S_b^2$ partition function of certain surface operators, 
however the challenge would be to identify what the correct world volume theories are. 
It would also be interesting if we consider quiver generalization of the current setup, the corresponding D-brane construction can be found in \cite{Chen:2011sj}.
\paragraph{}
Various surface defects in 4d $\cN=2$ Superconformal QCD were studied extensively in \cite{Gomis:2014eya}, they are labeled by the representation of $A_{N_c-1}$ and the corresponding 2d $\cN=(2,2)$ world volume theories were also identified. In particular, the 2d $\cN=(2,2)$ SQCDA we studied in this work corresponds to the totally symmetric representation.  In the context of 4d $\cN=2$ superconformal index $\cI_{\rm \cN=2}$, which is a twisted partition function on $S^1\times S^3$, it was noted in \cite{Gaiotto:2012xa}, and further studied in \cite{Bullimore:2014nla}, \cite{Alday:2013kda}, \cite{Gadde:2013dda},
that the insertion of these surface defects can be implemented by the action of various difference operators acting on $\cI_{\cN=2}$. 
It would be interesting to investigate whether similar story can occur for inserting the surface defects into $\cZ_{S_b^4}$, 
that is when we couple 2d $\cN=(2,2)$ SQCDA to 4d $\cN=2$ SQCD,  
whether we can replace the residue evaluation in section \ref{Sec:4dEvaluation} by the action of certain shift operator now acting on the various parts in the integrand, including the non-perturbative instanton part.
\paragraph{}
We hope to report on these in a future publication.

\acknowledgments
This work was supported in part by National Science Council through the grant No.101-2112-M-002-027-MY3, Center for Theoretical Sciences at National Taiwan University and Kenda Foundation. We would like to thank various enlightening discussions with Kazuo Hosomchi during the completion of this work.
HYC would also like to thank the hospitality of KEK theory group, University of Tokyo, Kavli Institute for the Physics and the Mathematics of the Universe (KIPMU), and Okinawa Institute of Science and Technology (OIST) where parts of this work were presented. 

\appendix
\section{Spinor Conventions and Auxiliary Fields on $S_{b^2}^4$}\label{App:Spinors}
\paragraph{}
Now let us summarize the Killing spinor conventions and other useful identities mostly used in Section \ref{Sec:HiggsLoc}, following \cite{Hama:2012bg}, our index conventions are:
\bea
&& A, B = 1, 2, \quad\quad \text{$SU(2)_R$ indices.}\nn\\
&& \alpha, \beta = 1, 2, \quad\quad  \text {chiral spinor indices}\nn\\
&& \dal, \dbt = 1, 2, \quad\quad \text{anti-chiral spinor indices.}\nn\\
&& a, b =1,2,3,4, \quad\quad \text {${\mathbb R}^4$ coordinate indices}\nn\\
&& m, n = \varphi,\chi, \theta, \rho, \quad\quad \text{$S_{b^2}^4$ coordinate indices}\nn\\
&& \hI, \hI = 1, 2, \dots 2n, \quad\quad \text{$Sp(n)$ indices}.\nn 
\eea
The chiral and anti-chiral spinors transform respectively as $({\bf 2, 1})$ and $({\bf 1,2})$ under the first and second factor of $SU(2)\times SU(2)\simeq SO(4)$ 4d rotation group,
their generators are given by following $2\times 2$ matrices:
\bea\label{def:sig}
&&\sig^a = - i\btau^a,\quad \bsig^a = i\btau^a, \quad a=1,2,3\nn\\
&&\sig^4 = {\bf 1}, \quad \bsig^4 = {\bf 1}.
\eea
The index structures are given by $\sig^a \equiv (\sig^a)_{\alpha\dal}$ and $\bsig^a = (\bsig^a)^{\dal\alpha}$, 
and they are raised or lower by the anti-symmetric $SU(2)$-invariant tensors $\eps^{\al\bt}, \eps^{\dal\dbt}, \eps_{\al\bt, }\eps_{\dal\dbt}$ whose non-vanishing elements are
\be
\eps^{12} = -\eps^{21} = -\eps_{12} = \eps_{21}=1.
\ee
The index contraction convention is such that, undotted indices $\al, \bt$ are suppressed when contracted in up-left, down-right order, 
while similarly for dotted indices $\dal, \dbt$ when contracted in down-left, up-right order.
We also define the following combinations:
\be
\sig_{ab} \equiv \frac{1}{2}(\sig_a\bsig_b-\sig_b\bsig_a)_\al^{~\bt} , \quad  \bsig_{ab} \equiv \frac{1}{2}(\bsig_a\sig_b-\bsig_b\sig_a)_{~\dbt}^{\dal} ,
\ee
such that $\sig_{ab}= -\frac{1}{2}\eps_{abcd} \sig^{cd}$ (anti-self-dual) and $\bsig_{ab} = \frac{1}{2} \eps_{abcd}\bsig^{cd}$ (self-dual).

\section{Supersymmetry Transformations on $S_{b^2}^4$} \label{App:SUSYTrans}
\paragraph{}
Before we list the relevant supersymmetry transformations for various field contents and the resultant supersymmetric Lagrangians, 
let us write down for our purposes the explicit form of auxiliary fields solved in \cite{Hama:2012bg}:
\bea
&&(T^{mn}\sig_{mn})_\alpha^{~\beta} =+ \frac{i}{4 f g}
\begin{pmatrix}
0 & e^{-i\theta}[(g-f)-ih]\\
e^{i\theta}[(g-f)+i h]& 0
\end{pmatrix},\\
&&(\bar{T}^{mn}\bsig_{mn})_{~\dbt}^{\dal} = +\frac{i}{4 f g}
\begin{pmatrix}
0 & e^{-i\theta}[(g-f)+ih]\\
e^{i\theta}[(g-f)-i h]& 0
\end{pmatrix},\\
&&(S^{mn}\sig_{mn})_\alpha^{~\beta} = -\frac{i}{4 f g}
\begin{pmatrix}
0 & e^{-i\theta}[(g+f)-ih]\\
e^{i\theta}[(g+f)+i h]& 0
\end{pmatrix},\\
&&(\bar{S}^{mn}\bsig_{mn})_{~\dbt}^{\dal} = -\frac{i}{4 f g}
\begin{pmatrix}
0 & e^{-i\theta}[(g+f)+ih]\\
e^{i\theta}[(g+f)-i h]& 0
\end{pmatrix}.\\
&&{\bf M}=\frac{1}{f^2}-\frac{1}{g^2}+\frac{h^2}{f^2 g^2}-\frac{4}{fg}.
\eea
these expressions  enter the Killing equations defining $(\xi_A,\bxi_A)$. 
Moreover here we also list few useful identities which are used in the main text:
\bea
&&\xi^{\bt A} \xi_{A\al} = \frac{1}{2} \sin^2\frac{\rho}{2} \delta^\bt_{~\al}, \quad \bxi_{\dal}^{~A} \bxi_A^{~\dbt} = -\frac{1}{2}\cos^2\frac{\rho}{2} \delta_{\dal}^{~\dbt},\label{KS-ID1}\\
&& \bxi_{\dal}^{~A}\xi_{A\al} = \xi^{\al A}\bxi_A^{~\dal} = -\frac{i}{2}\sin\frac{\rho}{2}\cos\frac{\rho}{2} (\cos\theta\tau^1 - \sin\theta \tau^2),\label{KS-ID2}\\
&& \bxi^{\dal}_{~A}\xi^{A\al} = \xi_{\al A}\bxi^A_{~\dal} = -\frac{i}{2}\sin\frac{\rho}{2}\cos\frac{\rho}{2} (\cos\theta\tau^1 + \sin\theta \tau^2),\label{KS-ID3}
\eea
note that $i\tau^{1,2} = -(\sig^{1,2})_{\al\dal} = (\bsig^{1,2})^{\dal\al}$, so that spinor index structures in the last two lines above still hold even though we suppress them here.
In the below we summarize the supersymmetric transformations of various field contents on $S^4_{b^2}$ used in main text,
we again follow the conventions in \cite{Hama:2012bg}.
\paragraph{}
{\bf Vector Multiplet}: A 4d $\cN=2$ vector multiplet contains a gauge field $A_m$, a pair of gauginos $\lam_{\al A}$ and $\blam_{~A}^{\dal}$, 
two scalar fields $\phi,\bphi$ and a symmetric auxiliary field $D_{AB}=D_{BA}$, all transform in the adjoint representation of the gauge group.
Their supersymmetry transformations on $S_{b^2}^4$ are parameterized by the killing spinors as $(\xi_A, \bxi_A)$:
\be
\label{VecTrans1}
\bQ A_m = i\xi^A\sig_m\blam_A-i\bxi^A\bsig_m\lam_A,
\ee
\be\label{VecTrans2}
\bQ \phi = -i\xi^A\lam_A,\quad \bQ\bphi = i\bxi^A\blam_A
\ee
\be\label{GauginoTrans1}
\bQ\lam_A=\frac{1}{2}\sig^{mn}\xi_A(F_{mn}+8\bphi T_{mn}-8\phi S_{mn})+2\sig^m\bxi_A D_m\phi+2i\xi_A[\phi,\bphi]+D_{AB}\xi^B,\\
\ee
\be \label{GauginoTrans2}
\bQ\blam_A=\frac{1}{2}\bsig^{mn}\bxi_A(F_{mn}+8\phi \bar{T}_{mn}-8\bar{\phi}\bar{S}_{mn})+2\bsig^m\xi_A D_m\bphi-2i\bxi_A[\phi,\bphi]+D_{AB}\bxi^B.
\ee
\bea
\bQ D_{AB}=&& -i\bxi_A\bsig^mD_m\lam_B-i\bxi_B\bsig^m D_m\lam_A+i\xi_A\sig^m D_m\blam_B+i\xi_B \sig^m D_m\blam_A\nn\\
&& -2[\phi, \bxi_A\blam_B+\bxi_B\blam_A]+2[\bphi, \xi_A\lam_B+\xi_B\lam_A].
\eea
The supersymmetric Lagrangian that closed under the off-shell supercharge $\bQ$ is then given by:
\bea\label{Lvec}
&&\cL_{\rm vec.} = {\rm Tr}{\Big[}\frac{1}{2}F_{mn}F^{mn}+16 F_{mn}(\bphi T^{mn}+\phi\bar{T}^{mn})+64\bphi^2 T_{mn}T^{mn}+64\phi^2\bar{T}_{mn}\bar{T}^{mn}\nn\\
&&-4D_m\bphi D^m \phi+2{\rm M}\bphi\phi-2i\lam^{A}\sig^m D_m \blam_A-2\lam^{A}[\bphi,\lam_A]+2\blam^A[\phi, \blam_A]
+4[\phi,\bphi]^2-\frac{1}{2} D^{AB}D_{AB}{\Big ]}\nn\\
\eea
To be compatible with the reality condition for $(\xi_A,\bxi_A)$ (\ref{KSReality}), various fields in the vector multiplet need to satisfy:
\be\label{RealityCond-Vec1}
A_m^{\dag} = A_m, \quad (\lam_{\al A})^{\dag} = \lam^{\al A}, \quad (\blam_{\dal A})^{\dag} = \lam^{\dal A}.
\ee
However for the off-shell vector multiplet Lagrangian (\ref{Lvec}) to be positive definite, when we perform the path-integration 
we need to deform the integration contour and replace the reality conditions for $\phi, \bphi$  and $D^{AB}$ with the following condition:
\be\label{RealityCond-Vec2}
\phi^{\dag} = -\bar{\phi}, \quad (D_{AB})^{\dag} = -D^{AB}.
\ee
Finally if the gauge group contains a $U(1)$ factor, we can have the following additional Lagrangian:
\be\label{LFI}
\cL_{\rm FI}=w^{AB} D_{AB} -{\rm M}(\phi+\bphi)-64\phi T^{kl}T_{kl}-64\bphi \bar{T}^{kl}\bar{T}_{kl}-8F^{kl}(T_{kl}+\bar{T}_{kl}),
\ee
where $w_{AB}=w_{BA}$ is given by:
\be\label{Def:wAB}
w_{AB}= \frac{4\xi_A\sig^{mn}\xi_B(T_{mn}-S_{mn})}{\xi^C\xi_C}=\frac{4\bxi_A\bsig^{mn}\bxi_B(\bar{T}_{mn}-\bar{S}_{mn})}{\bxi^C\bxi_C}.
\ee
Notice however due to the additional Higgs deformation term (\ref{HiggsDeform}) in the main text, we need to further relax reality condition for $D_{AB}$ and $\phi$ to ensure the positive definiteness and existence Higgs branch loci.
\paragraph{}
{\bf Hypermultiplet}: A set of $n$ $\cN=2$ hypermultiplet contains scalars $q_{A \hI}$ and pairs of fermions $(\psi_{\al \hI}, \bpsi^{\dal}_{\hI})$, 
and we also include auxiliary scalar fields $F_{A \hI}$. Here $\hI=1, 2, \dots, 2n$ are the $Sp(n)$ global symmetry group index. 
Suppressing the $Sp(n)$ indices, their supersymmetry transformations are given by:
\bea
\label{HyperTrans1}
\bQ q_A &=& -i\xi_A\psi + i\bxi_A\bpsi,\\
\label{HyperTrans2}
\bQ\psi &=& 2\sig^m\bxi_A D_m q^A + \sig^m D_m \bxi_A q^A-4i\xi_A(\bphi+\bar{\mu}) q^A+2\check{\xi}_A F^{A},\\
\label{HyperTrans3}
\bQ\bpsi &=& 2\bsig^m\xi_A D_m q^A + \bsig^m D_m \xi_A q^A-4i\bxi_A(\phi+\mu) q^A+2\bar{\check{\xi}}_A F^A,\\
\bQ F_A &=& +i\check{\xi}_A \sig^m D_m\bpsi - 2\check{\xi}_A (\phi+{\mu})\psi-2\check{\xi}_A\lam_B q^B+2i\check{\xi}_A(\sig^{kl}T_{kl})\psi\nn\\  
&& -i\bar{\check{\xi}}_A \bsig^m D_m\psi + 2\bar{\check{\xi}}_A(\bphi+\bar{\mu})\bpsi+2\bar{\check{\xi}}_A\blam_B q^B-2i\bar{\check{\xi}}_A(\bsig^{kl}\bar{T}_{kl})\bpsi,
\eea
where $(\check{\xi}_A,\bar{\check{\xi}}_A)$ are related to $(\xi_A, \bxi_A)$ via:
\be\label{Def: Checkxi}
\check{\xi}_{\al A} = \cot\frac{\rho}{2} \xi_{\al A}, \quad \bar{\check{\xi}}^{\dal}_{~A} = - \tan\frac{\rho}{2}\bar{\xi}^{\dal}_{~A}.
\ee
Notice that comparing with the supersymmetric transformation rules given in \cite{Hama:2012bg}, 
we have added the additional mass parameters $(\mu,\bar{\mu})$ for the hypermultiplets, their presence are crucial to ensure the existence of extra ``Higgs branch'' solutions to the saddle point equations.
We again need to impose on various fields the following reality constraints:
\bea
&& q_{\hI A}^{\dag} = q^{A\hI} = \Omega^{\hI\hJ}\eps^{AB} q_{\hJ B},\nn\\
&& \psi_{\al \hI}^{\dag} = \psi^{\al \hI} = \eps^{\al\bt} \Omega^{\hI\hJ} \psi_{\bt \hJ},\quad \bpsi_{\dal \hI}^{\dag} = \bpsi^{\dal \hI} = \eps^{\dal\dbt} \Omega^{\hI\hJ} \bpsi_{\dbt \hJ}\nn\\
&& F_{\hI A}^{\dag} = F^{A\hI} = \Omega^{\hI\hJ}\eps^{AB} F_{\hJ B}.
\eea
where $\Omega^{\hI\hJ}$ are the real $Sp(n)$ invariant rank two anti-symmetric tensor satisfying:
\be
\Omega_{\hI\hJ} = -\Omega_{\hJ\hI}, \quad (\Omega^{\hI\hJ})^{*} = -\Omega_{\hI\hJ}, \quad \Omega^{\hI\hK}\Omega_{\hK \hJ} = \delta^{\hI}_{\hJ}.
\ee
The covariant derivatives acting on the scalar and fermion $(q_A, \psi_\al)$ are defined to be:
\bea
D_n q_{\hI A} &=& \partial_n q_{\hI A} - i(A_n)_{\hI}^{~\hJ} q_{\hJ A}+i q_{\hI B}(V_n)^B_{~A}\,\label{Def:Dq}\\
D_n \psi_{\al\hI} &=& \partial_n \psi_{\al\hI} - i(A_n)_{\hI}^{~\hJ}\psi_{\al \hJ}+\frac{1}{4}\Omega^{ab}_n(\sig_{ab})_\al^{~\bt}\psi_{\bt\hI},\label{Def:Dpsi}
\eea
where $(A_n)_{\hI}^{\hJ}$ represent the coupling with the gauge field in the vector multiplet and $(V_m)^B_{~A}$ is the background $SU(2)_R$ gauge field such that $q_{\hI A}$ transform as doublet.
We have not given explicit form of $V_n$, however it is important to note they are non-vanishing.
The supersymmetric Lagrangian for the hypermultiplet where we have suppressed the $Sp(n)$ indices $\hI, \hJ$ is given by:
\bea\label{Lhyp}
\cL_{\rm hyp.} &=& \frac{1}{2} D_m q^A D^m q_A - q^A \{\phi+\mu, \bphi+\bar{\mu}\}q_A+\frac{i}{2} q^A D_{AB}q^B+\frac{1}{8}({\rm R}+{\rm M}) q^A q_A -\frac{i}{2}\bpsi\bsig^m D_m \psi\nn\\
&-&\frac{1}{2}\psi(\phi+\mu)\psi+\frac{1}{2} \bpsi (\bphi+\bar{\mu})\bpsi +\frac{i}{2}\psi\sig^{kl}T_{kl}\psi-\frac{i}{2}\bpsi\bsig^{kl}\bar{T}_{kl}\bpsi
-q^A \lam_A \psi + \bpsi\blam_A q^A-\frac{1}{2} F^A F_A,\nn\\
\eea
and to ensure the off-shell action to be positive definite, we again need to deform the integration contour for the auxiliary field $F_{\hI A}$ to impose that:
\be\label{Reality: F}
(F_{\hI A})^{\dag} = - F^{A\hI}. 
\ee
It was noticed in \cite{Hama:2012bg} that (\ref{Lhyp}) is in fact $\bQ$ exact, such that $\cL^{\rm hyp.} = \bQ \cV^{\rm hyp.}$.

\section{The instanton partition function with $U(1)$ gauge group}\label{U(1)}
\paragraph{}

Here we do a simple test by considering the case with $U(1)$ gauge group shown in \cite{Pan:2015hza}. Following the same splitting procedure on the instanton partition function in the section \ref{Sec:4dEvaluation}
and extracting the $Y$-independent term from instanton partition function. 
Because $Z^{\rm vec}_{ab}$ is independent of the vev of scalar,
such a term only comes from the residue of $Z^{\rm hyp}_{aI}$ which cancel with the remaining term from the residue of one loop hyper-multiplet determinant. So we are left with 
\be
\bZ^{U(1)}_{\{m,n\}}=\sum_{Y} \bZ^{U(1)}_{2d}\times\bZ^{U(1)}_{\rm extra}
\ee
\bea\label{ZU(1)2d}
\bZ^{U(1)}_{2d}&=&q^{|{\bY_{2d}}|}\fc{(-1)^{|\bY_{2d}|}\cdot\pd^{m}_{s=1}(1+b^{2}s-Y_{1}+Y_{s})_{Y_{1}}\pd^{m}_{s=1}(1+b^2(s-m)+Y_{s}-Y_{m+1})^{-1}_{Y_{m+1}}}{\pd^{m}_{s=1}(1+b^{2}s)_{Y_{s}}\pd^{m}_{r,s=1}(1+b^{2}(s-r)+Y_{s}-Y_{r})_{Y_{r}-Y_{r+1}}\pd^{m}_{r=1}(b^{2}(r-m-1))_{Y_{r}}}\nb\\\nb\\
&\times&\pd^{N_{f}}_{I=1}\pd^{m}_{r=1}\fc{(i(\hmu_{I}-\hmu_{l})b+b^2(r-m-1))_{Y_{r}}}{(i(\hmu_{I}-\hmu_{l})b+b^2(r-m-1)+Y_{r}-n)_{n}}
\eea
\bea\label{ZU(1)extra}
&&\bZ^{U(1)}_{\rm extra}=(-q)^{|{Y}/{\oY}|}\fc{1}{\pd^{Y^{\vee}_{1}}_{r,s=m+1}(1+b^{2}(s-r)+Y_{s}-Y_{r})_{Y_{r}-Y_{r+1}}}\times\fc{1}{\pd^{m}_{r=1}\pd^{Y^{\vee}_{1}}_{s=m+1}(1+b^2(s-r)+Y_{s}-Y_{r})_{Y_{r}-Y_{r+1}}}\nb\\\nb\\
&&\times\fc{1}{\pd^{Y^{\vee}_{1}}_{r=m+1}\pd^{m}_{s=1}(1+b^2(s-r)+Y_{s}-Y_{r})_{Y_{r}-Y_{r+1}}}\times\fc{1}{\pd^{Y^{\vee}_{1}}_{r=m+1}(b^{2}(r-Y^{\vee}_{1}-1))_{Y_{r}}}\nb\\\nb\\
&&\times\pd^{Y^{\vee}_{1}}_{s=m+1}\fc{(1+b^{2}s-Y_{1}+Y_{s})_{Y_{1}}}{(1+b^{2}s)_{Y_{s}}}\times\pd^{m}_{r=1}\fc{(b^2(r-m-1))_{Y_{r}}}{(b^2(r-Y^{\vee}_{1}-1))_{Y_{r}}}\times\pd^{m}_{s=1}(-1)^{Y_{m+1}}(b^2(m-s)-Y_{s})_{Y_{m+1}}\nb\\\nb\\
&&\times\pd^{N_{f}}_{I=1}\pd^{Y^{\vee}_{1}}_{r=m+1}(i(\hmu_{I}-\hmu_{l})b+b^2(r-m-1)-n)_{Y_{r}}
\eea
For comparison with above equations, we rewrite the $\bZ^{\rm inst.}_{\rm extra}$ (\ref{Zinstextra}) in the following form:
\bea\label{Zinstextra2}
\bZ^{\rm inst.}_{\rm extra}&=&q^{|\vec{Y}/\vec{\oY}|}(-1)^{N_{c}|\vec{Y}/\vec{\oY}|}\pd^{N_{c}}_{\ra,\rb=1}\Bigg{\{}\pd^{Y^{\vee}_{\ra 1}}_{r=m_{\ra}+1}\pd^{Y^{\vee}_{\rb 1}}_{s=m_{\rb}+1}\fc{1}{\left(1-x+b^2(s-r)+Y_{\rb s}-Y_{\ra r}\right)_{Y_{\ra r}-Y_{\ra (r+1)}}}\nb\\\nb\\
&\times&\pd^{m_{\ra}}_{r=1}\pd^{Y^{\vee}_{\rb 1}}_{s=m_{\rb}+1}\fc{1}{\left(1-x+b^2(s-r)+Y_{\rb s}-Y_{\ra r}\right)_{Y_{\ra r}-Y_{\ra (r+1)}}}\nb\\\nb\\
&\times&\pd^{Y^{\vee}_{\ra 1}}_{r=m_{\ra}+1}\pd^{m_{\rb}}_{s=1}\fc{(-1)^{Y_{\ra r}-Y_{\ra (r+1)}}}{\left(1-x+b^2(s-r)+Y_{\rb s}-Y_{\ra r}\right)_{Y_{\ra r}-Y_{\ra (r+1)}}}\nb\\\nb\\
&\times&\pd^{Y^{\vee}_{\ra 1}}_{r=m_{\ra}+1}\fc{1}{\left(x+b^2(r-Y^{\vee}_{\rb 1}-1)\right)_{Y_{\ra r}}}\times\pd^{Y^{\vee}_{\rb 1}}_{s=m_{\rb}+1}\fc{\left(1-x+sb^2+Y_{\rb s}-Y_{\ra 1}\right)_{Y_{\ra 1}}}{\left(1-x+sb^2\right)_{Y_{\rb s}}}\nb\\\nb\\
&\times&\pd^{m_{\ra}}_{r=1}\fc{\left(x+b^2(r-m_{\rb}-1)\right)_{Y_{\ra r}}}{\left(x+b^2(r-Y^{\vee}_{\rb 1}-1)\right)_{Y_{\ra r}}}\times\pd^{m_{\rb}}_{s=1}\left(x+b^2(m_{\ra}-s)-Y_{\rb s}\right)_{Y_{\ra(m_{\ra}+1)}}\Bigg{\}}\nb\\\nb\\
&\times&\pd^{N_{c}}_{\ra=1}\pd^{N_{f}}_{I=1}\pd^{Y^{\vee}_{\ra 1}}_{r=m_{\ra}+1}\left(i(\hmu_{I}-\hmu_{l_{\ra}})b+b^2(r-m_{\ra}-1)-n_{\ra}\right)_{Y_{\ra r}}
\eea
where the $x=i(\hmu_{l_{\rb}}-\hmu_{l_{\ra}})b-(m_{\ra}-m_{\rb})b^2-(n_{\ra}-n_{\rb})$ for $U(N_{c})$ gauge group and reduce to zero for $N_{c}=1$. 
By comparing the equations (\ref{Zinst<m}) and (\ref{Zinstextra2}) with (\ref{ZU(1)2d}) and (\ref{ZU(1)extra}) term by term, we show that the general expressions of (\ref{Zinst<m}) and (\ref{Zinstextra}) are indeed reduced to $U(1)$ results.

\bibliographystyle{sort}

\end{document}